\documentclass[a4paper,11pt]{article}
\pdfoutput=1
\usepackage{jcappub}
\usepackage{graphicx}
\usepackage{amssymb}
\usepackage{slashed}
\newcommand{\be}{\begin{eqnarray}}
\newcommand{\ee}{\end{eqnarray}}

\newcommand{\nn}{\nonumber}

\usepackage{amsmath,amssymb}
\usepackage{mathtools}
\usepackage{float}
\usepackage{color}
\usepackage{booktabs}
\usepackage{subcaption}
\usepackage{amsfonts}
\usepackage{multirow}
\title{Reconciling low multipole anomalies and reheating in single field inflationary 
models}
\author[a,1]{Rajesh Goswami,\note{Corresponding author.}}
\author[a]{ Urjit A. Yajnik}
\affiliation[a]{ Department of Physics,
Indian Institute of Technology Bombay, Mumbai-400076,
India.}
\emailAdd{rajesh@phy.iitb.ac.in}
\emailAdd{yajnik@phy.iitb.ac.in}

\abstract{
Reheating phase of inflationary Universe can be modeled by parameters $T_{\text{reh}}$, 
$\bar{w}_{\text{reh}}$ and $N_{\text{reh}}$, which can be constrained by the scalar spectral amplitude 
$A_{s}$ and the scalar spectral index $n_{s}$. On the other hand the low multipole anomalies in the CMB can 
be modeled by suitable features in the inflaton potential. We show that the parameters of these features in 
the inflaton potential provide additional constraints on the reheating parameters. For several single field 
models we find that the reheating parameters are substantially more constrained by the requirement of 
compatibility with the  proposed explanation for low multipole anomalies.
}
\begin{document}
\maketitle
\flushbottom
\section{Introduction}\label{sec1}
Slow-roll inflation 
\cite{Starobinsky:1980te,Guth:1980zm,Linde:1981mu,Hawking:1982cz,Linde:1983gd,Linde:2007fr,Martin:2013tda}
 predicts nearly scale invariant primordial power spectrum 
 \cite{Guth:1982ec,Starobinsky:1982ee,Mukhanov:1990me,Bardeen:1983qw,Riotto:2002yw} that provides a good fit 
to the recent cosmic microwave background (CMB) data 
\cite{Dunkley:2008ie,Komatsu:2008hk,0067-0049-192-2-18,0067-0049-208-2-19,Peiris:2003ff,0067-0049-192-2-16,
0004-637X-705-1-978, 
Reichardt:2008ay,Ade:2013zuv,Planck:2013jfk,Ade:2015xua,Ade:2015lrj}.
The inflationary phase terminates by \textquoteleft reheating\textquoteright~
phase  \cite{Turner:1983he,Traschen:1990sw,Albrecht:1982mp,Kofman:1994rk,Kofman:1997yn,Drewes:2013iaa,
Allahverdi:2010xz} so that the Universe  subsequently evolves according to the hot big-bang model 
\cite{Martin:2014nya,Munoz:2014eqa}. The physics of reheating phase is completely undetermined and 
 there are no direct observational data available, so far. However, indirect bounds can be obtained for the 
temperature  at the end of the
reheating ($T_{\text{reh}}$), the equation of state parameter, $w$,  during  reheating ($w_{\text{reh}}$) and 
the 
duration  of reheating ($N_{\text{reh}}$) from recent CMB data. 

Since slow roll inflation producing nearly scale invariant perturbations can be considered to have 
terminated if  $w$ attains 
$w> -\frac{1}{3}$,  it is convenient to assume that $w_{\text{reh}}$ = $-\frac{1}{3}$ at the 
beginning of the reheating.
After the reheating era, $w$ is expected to be  $\frac{1}{3}$, in order to make the Universe 
radiation dominated.
Nevertheless, the allowed range of $w$ during the reheating is considered to be $-\frac{1}{3}\leq 
w_{\text{reh}}\leq1$ in various scenarios. On the other hand, 
the lower and upper bounds of $T_{\text{reh}}$  are 10$^{-2}$ GeV (the big-bang nucleosynthesis (BBN)
temperature) \cite{Steigman:2007xt} and 10$^{16}$ GeV (inflation energy scale) 
respectively.\par

The relation between reheating and inflationary parameters can be obtained by considering the 
evolution of 
observable cosmological scales from the time of Hubble 
crossing during inflation to present time 
\cite{Liddle:2003as,Martin:2010kz,Dai:2014jja,Martin:2014nya}. 
For single-field inflationary models, we can derive relations among $T_{\text{reh}}$, 
$N_{\text{reh}}$, 
$w_{\text{reh}}$, the scalar power spectrum amplitude ($A_{s}$) and the scalar spectral index 
$n_{s}$. In 
addition,
CMB data is helpful in obtaining the constraints on $T_{\text{reh}}$ and $N_{\text{reh}}$  
\cite{Martin:2006rs,Martin:2010kz,Adshead:2010mc,Mielczarek:2010ag,Martin:2014nya,Dai:2014jja,
Cook:2015vqa}
.\par
The nearly scale invariant primordial power spectrum provides a good fit to the CMB data, and lends strong support to the
essential paradigm of inflation. However at lower multipoles, specifically around 
$\ell={\bf{22}}$ and $\bf{40}$, the data points lie outside the cosmic variance associated with the power law 
primordial spectrum. It indicates that certain non-trivial inflationary dynamics are possible.
If not of a completely accidental origin, the outlying data points could be important phenomenological inputs to
deducing ancillary details of the inflationary  model. There have been constant efforts 
\cite{Hannestad:2000pm,Bridle:2003sa,Mukherjee:2003ag,Hannestad:2003zs,Shafieloo:2003gf,Shafieloo:2006hs,
Shafieloo:2007tk,Nagata:2008zj,1475-7516-2009-07-011} to reconstruct the primordial power 
spectrum from the CMB anisotropies in a model independent way.
\par
The consideration of a burst of oscillations in the primordial power spectrum leads to a good fit to the 
CMB angular power spectrum, particularly around $\ell=\bf 22$ and 40 multipole 
moments \cite{Hazra:2010ve, Hazra:2016fkm,Hazra:2017joc}. In order to generate these oscillations in 
the primordial power spectrum, one has to consider a short 
period of deviation from slow-roll inflation \cite{Starobinsky:1992ts,Dvorkin:2009ne}. This deviation can be 
obtained by 
introducing a step in the inflaton 
potential \cite{Adams:2001vc,Peiris:2003ff,Covi:2006ci,Hamann:2007pa,Mortonson:2009qv,Joy:2007na,Jain:2008dw, 
Jain:2009pm}. A step with suitable height 
and width at a particular location of the inflationary potential has resulted in a better fit to the CMB data 
near 
the multipole $\ell={\bf 22}$.
In this paper we show that the generic relation between late time observables and reheating phase 
in a single field inflation can be strengthened by also demanding successful explanation of the low 
multipole anomalies. The link is the specific position of the inflaton in the course of its slow 
roll, $\phi_k/M_{Pl}$, at which it encounters the step in the potential. We obtain constraints on 
the reheating predictions of several inflationary
models by using the location of the step in the inflaton potential as obtained 
in ref. 
\cite{Hazra:2010ve}.  The investigation can be extended easily to other models, as will be done 
in a future publication.

The article is organized as follows:  Sec. \ref{sec2} deals with the slow-roll 
inflation and its 
predictions. 
In Sec. \ref{sec2}, we derive the expressions for $T_{\text{reh}}$ and $N_{\text{reh}}$ 
as functions of $\bar{w}_{\text{reh}}$ and the 
inflationary parameters ($\Delta N_{k}$ and $V_{\text{end}}$) to be described there. The expressions for 
$T_{\text{reh}}$ and 
$N_{\text{reh}}$ are derived as a function of the scalar spectral index $n_{s}$
for different single field inflationary 
models in Sec. \ref{sec3}. In addition, the observational bounds on 
inflationary models and reheating parameters are discussed 
for large field, small field (hilltop) and Starobinsky model by using Planck 2015 data 
\cite{Ade:2015xua,Ade:2015lrj}.
In Sec. \ref{sec4}, the effect of considering a step in the inflaton potential and its consequences on the 
constraints on
reheating parameters are detailed. Sec.\ref{sec5} contains the conclusions.

We work with $ \hbar=c=1$ units and the following values are used. $M_{\text{Pl}}=\sqrt\frac{1}{8\pi 
G} 
=2.435\times10^{18}\text{GeV}$ is the 
reduced Planck mass and  the redshift of last scattering surface is $z_{ls}=1100$. The $ z_{\text{eq}}=3365$ 
is the redshift of matter radiation 
equality and the present value of the Hubble parameter $ H_{0}=100 h$ km $\text{s}^{-1} \text{Mpc}^{-1} $ with
$h=0.68$ \cite{Ade:2015xua,Ade:2015lrj}.

\section{Reheating parameters in slow-roll models}\label{sec2}
We consider the inflaton $\phi$ governed by a potential $V(\phi)$ undergoing slow roll evolution 
with parameters $\epsilon$ and $\eta$, resulting in scalar curvature power spectrum $P_\zeta(k)$ and
tensor power spectrum $P_h(k)$ as a function of the Fourier transform variable $k$ of the argument 
of the spatial correlation functions, with corresponding indices $n_s-1$ and $n_T$. The details of 
the definitions and notation are standard \cite{Liddle:1993fq}, and can be found also in 
the references \cite{Riotto:2002yw,Bassett:2005xm,Martin:2013tda}.
We shall use $A_{s}$ and $A_{T}$, the amplitude of scalar and tensor power spectra at 
the pivot scale $k_{*}$ as used by Planck collaboration, $\frac{k_{*}}{a_{0}}=0.05 
\text{Mpc}^{-1}$. For  $k=k_{*}$, these amplitudes are given in terms of $H_{*}$ as
\be\label{e7}
A_{T}= P_{h}(k_{*})=\frac{2H_{*}^{2}}{\pi^{2}M_{\text{Pl}}^{2}}, \hspace{1cm} 
A_{s}= P_{\zeta}(k_{*})=\frac{H_{*}^{2}}{8\pi^{2}M_{\text{Pl}}^{2}\epsilon_{*}}.
\ee
In terms of the slow-roll parameters $\epsilon$ and $\eta$, the tensor to scalar ratio $r$, the 
scalar spectral index $n_{s}$ and the tensor spectral index 
$n_{T}$ satisfy the relations 
\be\label{ns}
 r=16\epsilon,\hspace{1cm} n_{s}=1-6\epsilon+2\eta,\hspace{1cm} n_{T}=-2\epsilon.
\ee
The total number of e-foldings, $N_{T}$, is defined as the logarithm of the ratio of the scale 
factor at the 
final 
time $t_{e}$ to it's value at initial time $t_{i}$ of the era of inflation.
\be
 N_{T}\equiv\ln\frac{a(t_{e})}{a(t_{i})}=\int_{t_{i}}^{t_{e}}{H 
dt}=\int_{\phi_{i}}^{\phi_{\text{end}}}{\frac{H}{\dot{\phi}}d\phi}=\frac{1}{M_{\text{Pl}}}\int^{\phi_{i}}_{
\phi_{\text{end}}}{\frac{1}{\sqrt{2\epsilon}}d\phi}.
\ee
Where $\phi_{i}$ and $\phi_{\text{end}}$ are the initial and final values of the inflaton field $\phi$ and 
$\epsilon$ is the slow-roll parameter defined as $\epsilon=-\dot{H}/H^{2}=\dot{\phi}^2/(2H^2 
M_\mathrm{Pl}^2 )$.
Likewise, given a mode $k$, the number of e-foldings between the time when it 
crosses the Hubble 
horizon and the end of inflation is given by
\be\label{NKE}
 \Delta 
N_{k}=\int_{\phi_{k}}^{\phi_{\text{end}}}{\frac{H}{\dot{\phi}}d\phi}=\frac{1}{M_{\text{Pl}}}\int^{\phi_{k}}_{
\phi_{\text{end}}}{\frac{1}{\sqrt{2\epsilon}}d\phi},
\ee
where $\phi_{k}$ is the value of the inflaton field at the time of Hubble crossing of the scale k. For the 
slow-roll approximation i.e., $V(\phi)\gg\dot{\phi}^{2}$ and 
$\ddot{\phi}\ll3H\dot{\phi}$ the eq. \eqref{NKE} 
becomes 
\be\label{e11}
 \Delta N_{k}\approx\frac{1}{M_{\text{Pl}}^{2}}\int_{\phi_{\text{end}}}^{\phi_{k}}{\frac{V}{V'}d\phi}.
\ee
We shall also be interested in the situation where the slow roll condition is briefly violated. This happens when 
the inflaton negotiates the step in the potential. However for a small enough step, $\ddot{\phi}$ becomes
appreciable only briefly and the kinetic energy  $\frac{1}{2}{\dot{\phi}}^2$ does not grow appreciably \cite{Hamann:2007pa,Adshead:2011jq}. 
The brief 
departure from slow roll can be
accounted for by an additional quantity $\Delta N_{\mathrm{step}}$ which can be shown to remain negligible
compared to the main value of interest $ \Delta N_{k}$. We demonstrate this in detail in Appendix \ref{Ap}.

We now turn to relating the observed wavenumber of any physical scale today $\frac{k}{a_{0}}$,
to its value  at the time of Hubble crossing during inflation   
$\frac{k}{a_{k}}$. This can be obtained as
\be\label{e12}
\frac{k}{a_{k}}&=&\frac{k}{a_{0}}\frac{a_{0}}{a_{k}},\nn\\
&=&\frac{k}{a_{0}}\frac{a_{0}}{a_{\text{eq}}}\frac{a_{\text{eq}}}{a_{\text{reh}}}\frac{a_{\text{reh}}}{a_{
\text{end}}}\frac{a_{\text{end}}}{a_{k}}.
\ee
Here $a_{k}$, $a_{\text{eq}}$ and $a_{0}$ represent the value of scale 
factor at the time of Hubble crossing,  matter radiation equality and at present time 
respectively. The somewhat ill defined but physically significant epochs  $a_{\text{end}}$, 
$a_{\text{reh}}$ represent end of inflation and the end of reheating respectively. For convenience 
one also introduces 
\be
 N_{\text{reh}}\equiv\ln\left(\frac{a_{\text{reh}}}{a_{\text{end}}}\right)\qquad \textrm{and} 
\qquad \Delta N_{k}\equiv \ln \left(\frac{a_{\text{end}}}{a_{k}}\right)
\ee
in line with the number of e-foldings. The dynamic quantity $\Delta N_{k}$ represents the 
number of e-folds remaining after the scale $k$ has crossed the Hubble radius during 
inflation.
The $a_{\text{reh}}$ demarkates successful return to a radiation dominated Universe. Hence the $N_{\text{reh}}$ encodes both, 
an epoch of preheating 
\cite{Boyanovsky:1996ks,Kofman:1997yn,Kofman:1997pt,Felder:1998vq,
Giudice:2001ep,Desroche:2005yt} as well as a subsequent thermalization 
process.
Subsequent evolution of the Universe governed by an energy density
\be\label{e24}
 \rho_{\text{reh}}=\frac{\pi^{2}}{30}g_{\text{reh}}T_{\text{reh}}^{4},
\ee
where $T_{\text{reh}}$ is the temperature, and $g_{\text{reh}}$ is the effective number of 
relativistic species at the end of reheating. We further consider the energy density at the 
end of reheating to be entirely radiation type only i.e., 
$\rho(T_\text{reh})=\rho^{\text{radiation}}(T_{\text{reh}})$.
Now, using $z_{\text{eq}}$ as the redshift at the epoch of matter-radiation equality,
eq. \eqref{e12}  can be recast as
\be\label{e14} 
\frac{k}{a_{k}}=\frac{k}{a_{0}}(1+z_{\text{eq}})\left(\frac{\rho_{\text{reh}}}{\rho_{\text{eq}}}\right)^{1/4}
e^{N_{\text{reh}}}e^{\Delta N_{k}}.
\ee
As a further parametrization of the  reheating phase, it is assumed to be
dominated by a fluid \cite{Martin:2003bt, Martin:2013tda}  of pressure P and energy density 
$\rho$, with equation of state $w_{\text{reh}}=\frac{P}{\rho}$. Imposing the continuity equation, we have
\be
\dot{\rho}+3H(\rho+P)=0,\\
\dot{\rho}+3H\rho(1+w_{\text{reh}})=0.
\ee
In view of this equation, we have
\begin{eqnarray}
\label{e17}
\rho_{\text{reh}}&=&\rho_{\text{end}} e^{- 3N_{\text{reh}}(1+\bar{w}_{\text{reh}})}; \\
\textrm{where} \qquad \bar{w}_{\text{reh}}=<w>&=&\frac{1}{ 
N_{\text{reh}}}\int_{N_{e}}^{N}{w_{\text{reh}}(N)}dN
\end{eqnarray}
Here $\bar{w}_{\text{reh}}$ is the 
average equation of state parameter during reheating \cite{Martin:2014nya}.
eq. \eqref{e17} can now be rewritten in the following form
\be\label{e18}
\frac{a_{\text{reh}}}{a_{\text{end}}}=e^{
N_{\text{reh}}}=\left(\frac{\rho_{\text{reh}}}{\rho_{\text{end}}}\right)^{-\frac{1}{3+3\bar{w}_{\text{reh}}}}.
\ee
Using eq. \eqref{e18} and eq. \eqref{e24} we can write the reheating e-folds 
$N_{\text{reh}}$ as
\be\label{e26}
N_{\text{reh}}=\frac{1}{3(1+\bar{w}_{\text{reh}})}\left\{\ln\left(\frac{3}{2}V_{\text{end}}
\right)-\ln\left(\frac{\pi^{2}}{30}g_{\text{reh}}\right)\right\}-\frac{4}{3(1+\bar{w}_{\text{reh}})}
\ln T_ {\text{reh}}.
\ee
Thus returning to the quantity \eqref{e12}, substituting eq. \eqref{e18} in eq. 
\eqref{e14} we 
have
\be\label{e19}
\frac{k}{a_{k}}=H_{k}=\frac{k}{a_{0}}(1+z_{\text{eq}})\left(\frac{\rho_{\text{reh}}}{\rho_{\text{eq}}}\right)^
{\frac{1}{4}}
\left(\frac{
\rho_{\text{reh}}}{\rho_{\text{end}}}\right)^{-\frac{1}{3+3\bar{w}_{\text{reh}}}}e^{\Delta N_{k}}.
\ee
To make a further contact with slow roll inflation, we begin by noting that the parameter 
$\epsilon$ has an expression valid without restriction to slow roll condition,
\be\label{eps}
\epsilon=-\frac{\dot{H}}{H^{2}}=\frac{\frac{3}{2}\dot{\phi^{2}}}{\frac{1}{2}\dot{\phi^{2}}+V(\phi)} .
\ee
From eq. \eqref{eps} we can write the kinetic energy of the inflaton field in 
terms of $\epsilon$ and 
$V(\phi)$ as given below
\be\label{kinetic}
\frac{1}{2}\dot{\phi}^{2}=\frac{\epsilon V(\phi)}{3-\epsilon}.
\ee
Now, using eq. \eqref{kinetic} the energy density of the Universe and the Hubble 
parameter during inflation  
can be written as
\be
\rho(\phi)&=&\frac { 1 } {2}\dot{\phi^{2}}+V(\phi)=\frac{3
V(\phi)}{3-\epsilon},\\
\label{e21}
H^{2}&=&\frac{1}{3M_{\text{Pl}}^{2}}\rho=\frac{1 }{M_{\text{Pl}}^{2}} 
\left(\frac{
V(\phi)}{3-\epsilon}\right).
\ee
In general, the  slow-roll parameter $\epsilon$ becomes equal to 1 at the  end of 
inflation. Hence, the energy density at the end of inflation is 
$\rho_{\text{end}}=\frac{3}{2}V_{\text{end}}$, where 
$V_{\text{end}}=V(\phi_{\text{end}})$ and $\phi_{\text{end}}$ respectively represent the 
potential and field at the end of inflation. In terms of these quantities,  eq. 
\eqref{e19} can be 
written as
\be\label{e20}
 \frac{k}{a_{k}}=H_{k}=\frac{k}{a_{0}}\left(1+z_{\text{eq}}\right) 
\rho_{\text{reh}}^{\frac{3\bar{w}_{\text{reh}}-1}{12(1+\bar{w}_{\text{reh}})}} 
\rho_{\text{eq}}^{-\frac{1}{4}}\left(\frac{3}{2}V_{\text{end}}\right)^{\frac{1}{3(1+\bar{w}_{\text{reh}})}} 
e^{\Delta N_{k}} .
\ee
 From eq. \eqref{e20}, $\Delta N_{k} $ can be obtained as
\be\label{e23}
 \Delta N_{k}=&\ln H_{k}-\ln(\frac{k}{a_{0}})-\ln 
(1+z_{\text{eq}})-\frac{3\bar{w}_{\text{reh}}-1}{3(1+\bar{w}_{\text{reh}})}\ln(\rho_{\text{reh}}^{
\frac{1}{4}}
)+\ln(\rho_{\text{eq}}^{\frac{1}{4}
} )\nn\\
&-\frac { 1}{ 3(1+\bar{w}_{\text{reh}})}\ln\left(\frac{3}{2}V_{\text{end}}\right) .
\ee
eq. \eqref{e23} can be inverted, and using eq. \eqref{e24} we can obtain a 
mutual consistency relation 
between the various parameters introduced,  
\be\label{e25}
 \ln T_{\text{reh}}=&\frac{3(1+\bar{w}_{\text{reh}})}{3\bar{w}_{\text{reh}}-1}\left\{\ln H_{k}-\ln 
\frac{k}{a_{0}}-\ln(1+z_{\text{eq}})-\Delta 
N_{k}+\ln(\rho_{\text{eq}}^{\frac{1}{4}})\right\}\nn\\
&-\frac{1}{3\bar{w}_{\text{reh}}-1}\ln\left(\frac{3}{2}V_{\text{end}}\right) 
-\frac{1}{4}\ln\left(\frac{\pi^{2}}{30}g_{\text{reh}}\right) .
\ee
substituting the expression for $T_{\text{reh}}$ from eq. \eqref{e26} in eq. 
\eqref{e25} we get the 
expression for $N_{\text{reh}}$ as below
\be\label{e27}
N_{\text{reh}}=\frac{1}{3\bar{w}_{\text{reh}}-1}\ln\left(\frac{3}{2}V_{\text{end}}\right)+\frac{4}{3\bar{w}_{
\text{
reh}}-1}&\Big{\{}\ln\left(\frac{k}{a_{0}}\right)+\ln\left(1+z_{\text{eq}}\right)+\Delta 
N_{k}\nn\\
&-\ln H_{k}-\ln\left(\rho_{\text{eq}}^{\frac{1}{4}}\right)\Big{\}} .
\ee
eq.s \eqref{e25} and \eqref{e27} are the key relationships we shall use for 
relating late time  
observables and internal consistency among the reheating parameters for two specific models.
It is reasonable to assume $g_{\text{reh}}\approx100$ for our calculations \cite{Dai:2014jja}.
\section{Models and constraints}\label{sec3}
\subsection{Large field quadratic model}
\begin{figure}
\begin{subfigure}{.47\textwidth}
 \centering
  \includegraphics[width=1\linewidth]{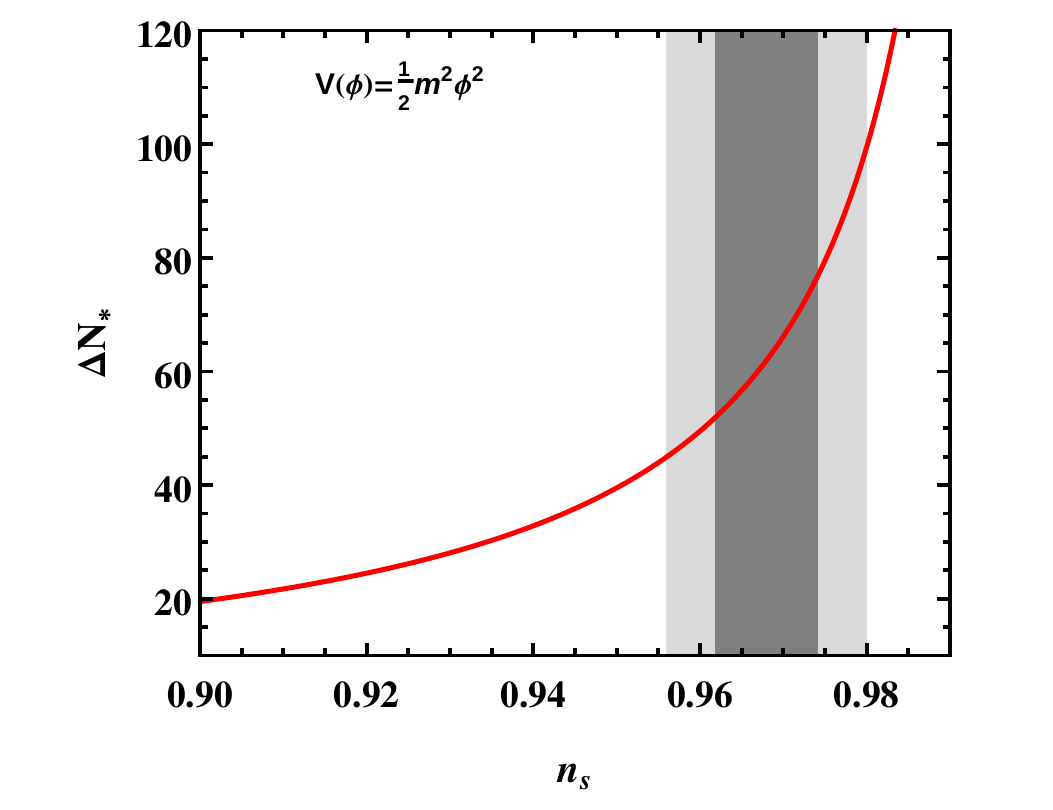}
  \caption{}
  \label{fig1a}
\end{subfigure}
\begin{subfigure}{.47\textwidth}
  \centering
  \includegraphics[width=1\linewidth]{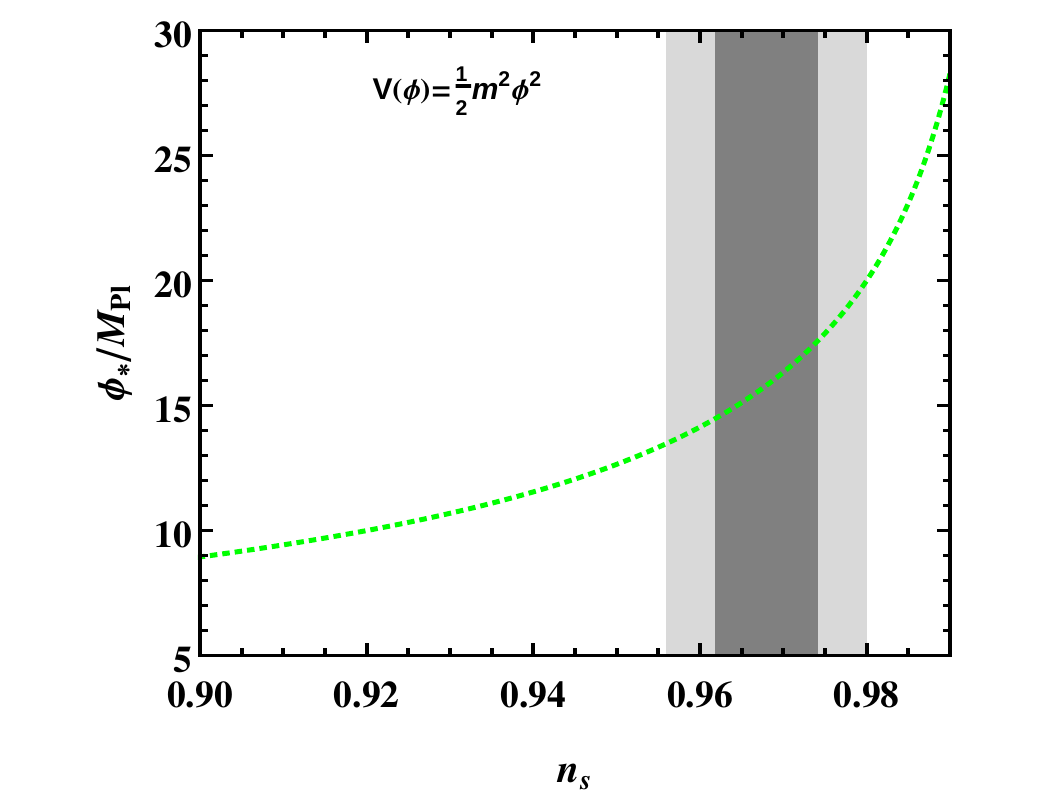}
  \caption{}
  \label{fig1b}
\end{subfigure}
\caption{The variation of (a)  $\Delta N_{*}$  and (b) $\frac{\phi_{*}}{M_{\text{Pl}}}$  as a function of 
scalar 
spectral index $n_{s}$ within the large field model $V(\phi)=\frac{1}{2}m^{2}\phi^{2}$.
The dark gray and light gray shaded regions
correspond to the 1$\sigma $ and 2$\sigma$ bounds respectively on $n_{s}$ from Planck 2015 data (TT+Low P + 
Lensing) 
\cite{Ade:2015xua,Ade:2015lrj}.}
\label{fig1}
\end{figure}

\begin{figure}
\begin{center}
\begin{subfigure}{.5\textwidth}
  \includegraphics[width=1\linewidth]{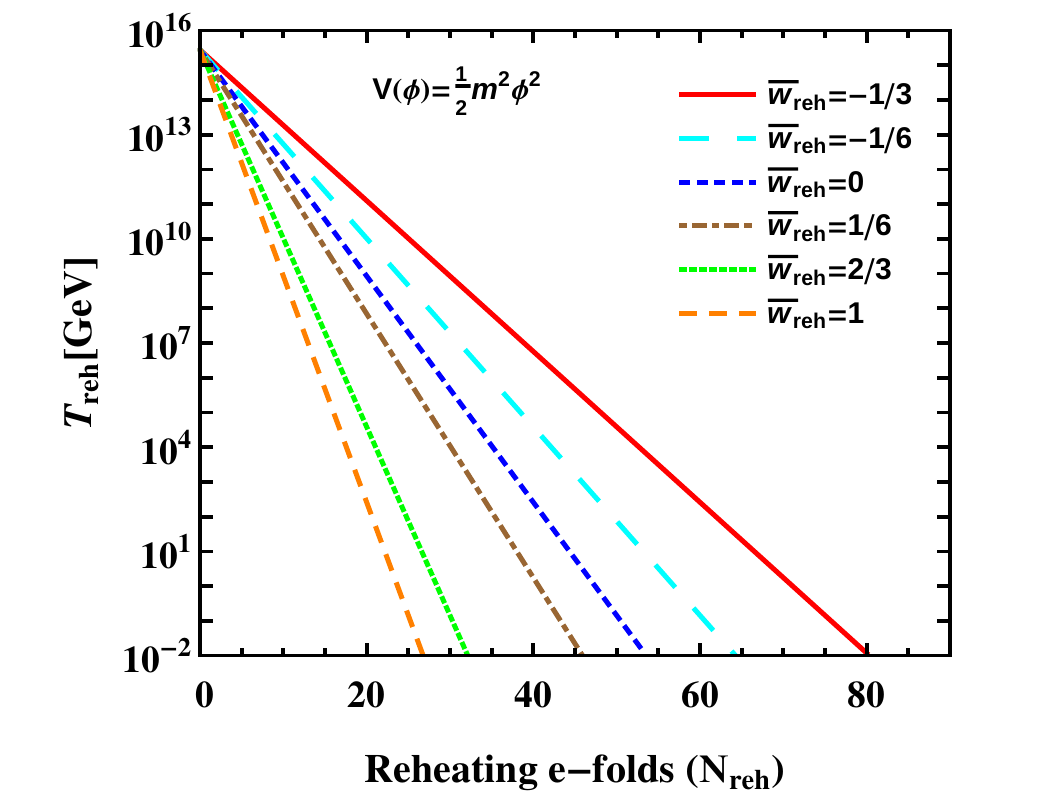}
  \caption{}
  \label{fig2(a)}
\end{subfigure}%
\begin{subfigure}{.5\textwidth}
  \includegraphics[width=1\linewidth]{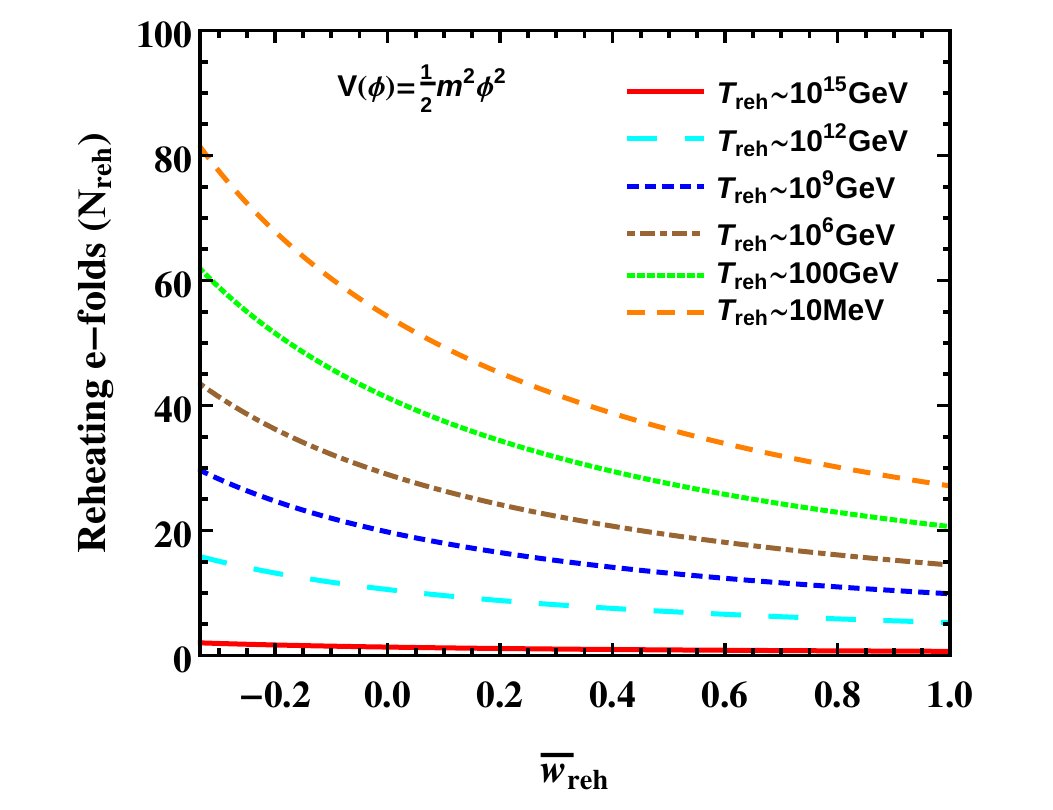}
  \caption{}
  \label{fig2(b)}
\end{subfigure}
\end{center}
\caption{Plots for Large field model ($V(\phi)=\frac{1}{2}m^{2}\phi^{2}$). Plot (a) shows the variation of  
$T_{\text{reh}}$ with $N_{\text{reh}}$ for different values of $\bar{w}_{\text{reh}}$: 
$\bar{w}_{\text{reh}}=-\frac{1}{3}$ (solid red), $\bar{w}_{\text{reh}}=-\frac{1}{6}$ (large dashed cyan), 
$\bar{w}_{\text{reh}}=0$ (small dashed blue), $\bar{w}_{\text{reh}}=\frac{1}{6}$ (dot-dashed brown), 
$\bar{w}_{\text{reh}}=\frac{2}{3}$ (tiny dashed green), $\bar{w}_{\text{reh}}=1$ (medium dashed orange). Plot 
(b) shows $N_{\text{reh}}$ as a function of $\bar{w}_{\text{reh}}$ for different values of $T_{\text{reh}}$: 
$T_{\text{reh}}\sim 10^{15} \text{GeV}$ (solid red), $T_{\text{reh}}\sim 10^{12} \text{GeV}$ 
(large dashed cyan),$T_{\text{reh}}\sim 10^{9} \text{GeV}$ (small dashed blue), $T_{\text{reh}}\sim 10^{6} 
\text{GeV}$ (dot-dashed brown), $T_{\text{reh}}\sim 100 \text{GeV}$ (tiny dashed green),  $T_{\text{reh}}\sim 
10\text{MeV} $ (medium dashed orange). }
\label{fig2}
\end{figure}

\begin{figure}
\begin{subfigure}{.5\textwidth}
 \centering
  \includegraphics[width=1\linewidth]{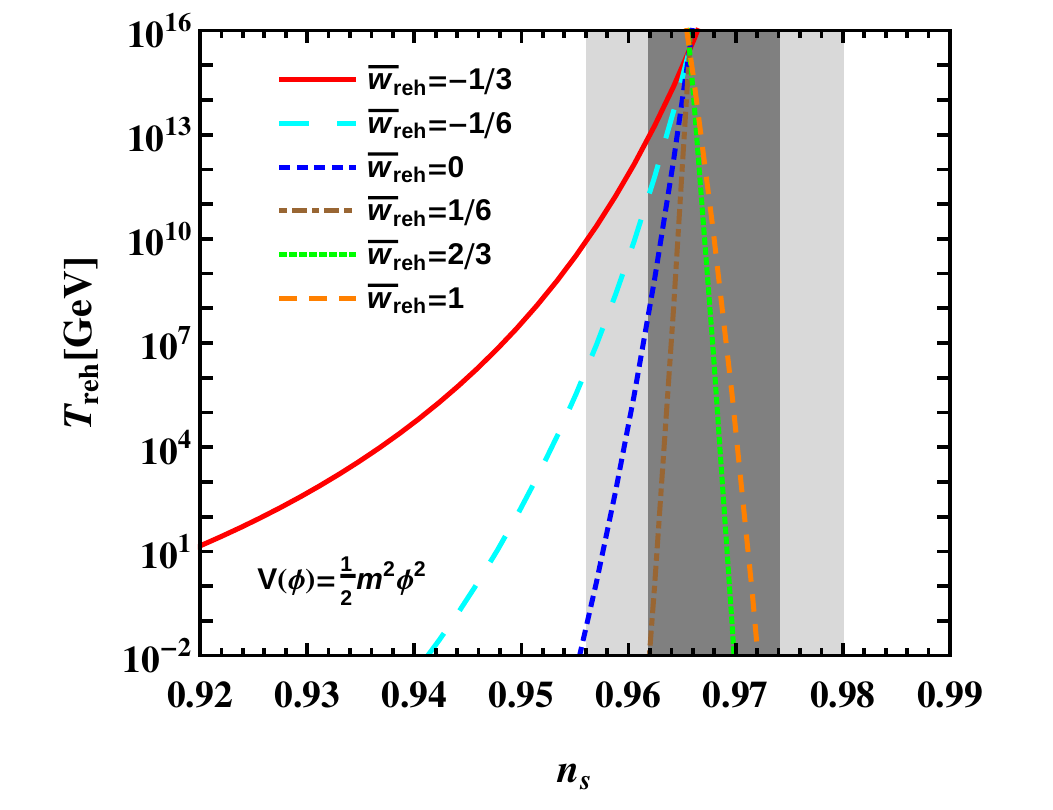}
  \caption{}
  \label{fig3a}
\end{subfigure}%
\begin{subfigure}{.5\textwidth}
  \centering
  \includegraphics[width=1\linewidth]{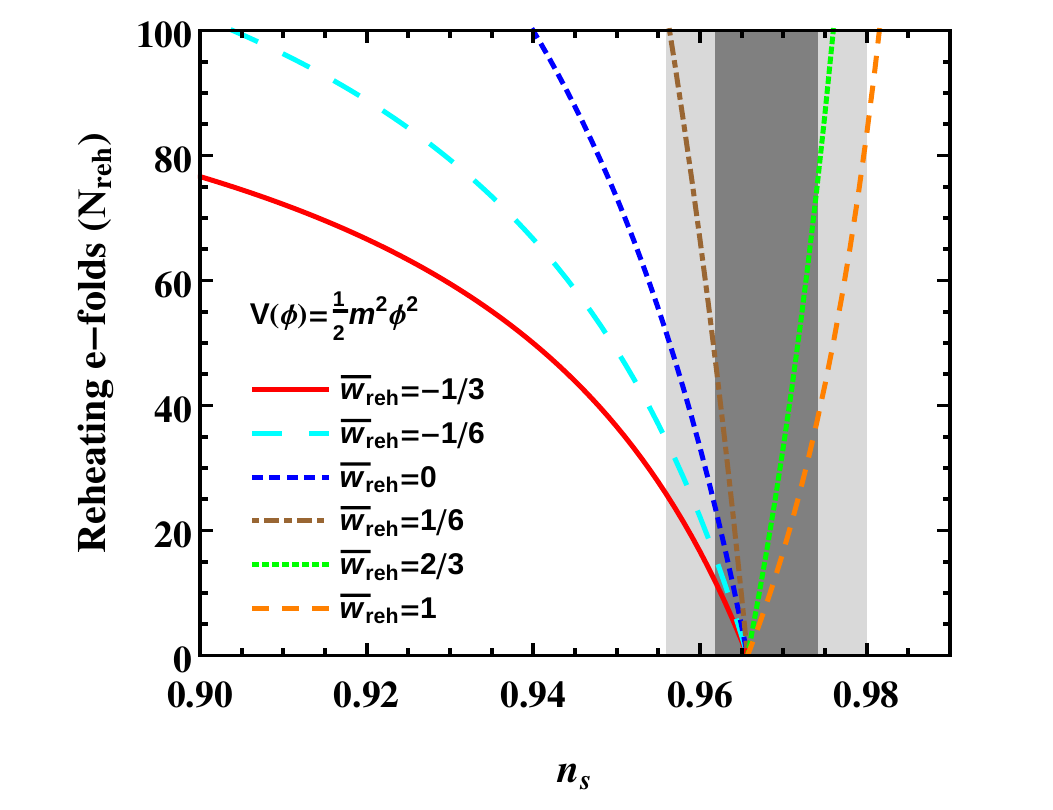}
  \caption{}
  \label{fig3b}
\end{subfigure}
\caption{Plots for Large field model ($V(\phi)=\frac{1}{2}m^{2}\phi^{2}$). (a) $T_{\text{reh}}$ and 
(b) $N_{\text{reh}}$ as a function of $n_{s}$ for different values of $\bar{w}_{\text{reh}}$:  
$\bar{w}_{\text{reh}}=-\frac{1}{3}$ 
(solid red),  
$\bar{w}_{\text{reh}}=-\frac{1}{6}$ (large dashed cyan),  $\bar{w}_{\text{reh}}=0$ (small dashed blue),  
$\bar{w}_{\text{reh}}=\frac{1}{6}$ (dot-dashed brown), 
  $\bar{w}_{\text{reh}}=\frac{2}{3}$ (tiny dashed green),  $\bar{w}_{\text{reh}}=1$ (medium dashed orange).
The dark gray and light gray shaded region corresponds to the 1$\sigma $ 
and 2$\sigma$ bounds on $n_{s}$ from Planck 2015 data (TT+Low P + Lensing) \cite{Ade:2015xua,Ade:2015lrj}.}
\label{fig3}
\end{figure}

Large field quadratic model \cite{Linde:1984st,Bassett:2005xm,Martin:2013tda,Dai:2014jja} of 
inflation is 
described by the potential 
$V(\phi)=\frac{1}{2}m^{2}\phi^{2}$.
For this model the Hubble parameter as defined in eq. \eqref{e21} at the time of 
Hubble 
crossing of  the 
scale $k$ takes the form
\be\label{Hubble}
H_{k}^{2}=\frac{1}{M_{\text{Pl}}^{2}}\left(\frac{V_{k}}{3-\epsilon_{k}}\right)=\frac{1}{M_{\text{Pl}}^{2}}
\Bigg{(}\frac{
	\frac{1}{2}m^{2}\phi_{k}^{2} } { 3-2\left(\frac{M_{\text{Pl}}}{\phi_{k}}\right)^{2} } \Bigg{ ) },
\ee
where $V_{k}$, $\phi_{k}$ and $\epsilon_{k}$ respectively represent the potential, inflaton field and the 
slow-roll 
parameter $\epsilon$ 
at the time of Hubble crossing of the mode $k$.
Now, consider the mode $k_{*}$ corresponding to the pivot scale introduced above \eqref{e7}, which 
crosses the Hubble radius $H_{*}$ during inflation when the field $\phi$ has attained the value 
$\phi_{*}$. The number of e-folds remaining after the pivot scale $k_{*}$ crosses the 
Hubble radius is
\be\label{e29}
\begin{aligned}
 \Delta N_{*}&\simeq\frac{1}{M_{\text{Pl}}^{2}}\int_{\phi_{\text{end}}}^{\phi_{*}}{\frac{V}{V'}d\phi}\\ 
&=\frac{1}{4}\Big{[}\left(\frac{\phi_{*}}{M_{\text{Pl}}}\right)^{2}-\left(\frac{\phi_{\text{end}}}{M_{\text{Pl
}}}\right)^{2}\Big{]}\\
&=\frac{1}{4}\Big{[}\left(\frac{\phi_{*}}{M_{\text{Pl}}}\right)^{2}-2\Big{]} .
 \end{aligned}
\ee
where we have used  the condition defining the end of inflation, $\epsilon=1$ which gives 
$\frac{\phi_{\text{end}}^{2}}{M_{\text{Pl}}^{2}}=2$.
Using $\epsilon=2M^2_{Pl}/\phi^2$ as arises in this model, the spectral index 
$n_{s}$, eq. 
\eqref{ns}, can be written as
\be\label{e32}
 n_{s}=1-8\left(\frac{M_{\text{Pl}}}{\phi_{*}}\right)^{2} .
\ee
And thus $\Delta 
N_{*}$ as a function of the scalar spectral index $n_{s}$ and is given by
\be
 \Delta N_{*}=\left(\frac{2}{1-n_{s}}-\frac{1}{2}\right) .
\ee
The variation of $\Delta N_{*}$ and $\frac{\phi_{*}}{M_{\text{Pl}}}$ 
with the scalar spectral index $n_{s}$ is shown in figure~{\ref{fig1}}. The dark gray and light 
gray shaded region corresponds to the 1$\sigma $ and 2$\sigma$ bounds on $n_{s}$ from Planck 2015 data 
(TT+Low 
P + Lensing) \cite{Ade:2015xua,Ade:2015lrj}.
Further, in this model one obtains the relation 
\be\label{e36}
 H_{*}=\pi M_{\text{Pl}}\sqrt{2A_{s}(1-n_{s})}.
\ee
where $n_s$ although strictly $k$ dependent has been replaced by it's almost 
constant value. This, 
along with the relation of $H$ and field $\phi$ in this model, and the criterion for the end of 
inflation as used in \eqref{e29}, gives the value of $V$ at
the end 
of the inflation, $V_{\text{end}}$, as a function of $A_{s}$ 
and $n_s$,
\be\label{e37}
V_{\text{end}}=\frac{1}{2}m^{2}\phi_{\text{end}}^{2}
\approx \frac{3}{2}\pi^{2}A_{s}M_{\text{Pl}}^{4}(1-n_{s})^{2} .
\ee
After substituting eq. \eqref{e36} and \eqref{e37} in eq. \eqref{e25} and  
\eqref{e27} we can write the 
temperature at the end of reheating, 
$T_{\text{reh}}$, and reheating 
e-folds, $ N_{\text{reh}}$, in terms of $A_{s}$ and $n_{s}$ for the pivot scale as given below 
\be\label{e38}
 \ln \left(T_{\text{reh}}\right)&=\frac{3(1+\bar{w}_{\text{reh}})}{3\bar{w}_{\text{reh}}-1}\Bigg{\{}\ln 
\left(\pi M_{\text{Pl}}\sqrt{2A_{s}(1-n_{s})}\right)-\ln 
\left(\frac{k_{*}}{a_{0}}\right)-\ln\left((1+z_{\text{eq}}\right)-\left(\frac{2}{1-n_{s}}-\frac{1}{2}
\right)\nn\\
&+\ln\left(\rho_{\text{eq}}^{\frac{1}{4}}\right)\Bigg{\}}
-\frac{1}{3\bar{w}_{\text{reh}}-1}\ln 
\left(\frac{9}{4}M_{\text{Pl}}^{4}\pi^{2}A_{s}(1-n_{s})^{2}\right)
-\frac {1} {4} \ln\left(\frac {\pi^{2}} {30}g_{\text{reh}}\right).
\ee
\be\label{e39}
N_{\text{reh}}=&\frac{1}{3\bar{w}_{\text{reh}}-1}\ln\left(\frac{9}{4}M_{\text{Pl}}^{4}\pi^{2}A_{s}(1-n_{s})^{2
}
\right)+\frac{4}{3\bar{w}_{\text{reh}}-1}\Bigg{\{}\ln\left(\frac{k_{*}}{a_{0}}\right)+\ln\left(1+z_{\text{eq}}
\right)\nn\\
& + \left(\frac{2}{1-n_{s}}-\frac{1}{2}\right)-\ln \left(\pi 
M_{\text{Pl}}\sqrt{2A_{s}(1-n_{s})}\right)-\ln\left(\rho_{\text{eq}}^{\frac{1
}{4}}\right)\Bigg{\}} .
\ee
Now, we need information about the mass scale ``$m$'' which can obtained by 
combining eqs. 
\eqref{Hubble} and \eqref{e36}. Using Planck's central value of 
$A_{s}=2.139 \times 
10^{-9} ~ \text{and}~ n_{s}=0.968$
\cite{Ade:2015xua,Ade:2015lrj} we obtain $m=1.38\times 10^{13}$ 
GeV and thus $V_{\text{end}}=(5.8\times 10^{15} \text{GeV})^{4}$. After substituting the values of 
$V_{\text{end}}$ and $g_{\text{reh}}\approx 100$ in eq. \eqref{e26}, one obtains 
the relation among 
$T_{\text{reh}}$, 
$N_{\text{reh}}$ and $\bar{w}_{\text{reh}}$ for this quadratic large field model as given below
\be\label{newN}
N_{\text{reh}}\approx \frac{142.12}{3(1+\bar{w}_{\text{reh}})}-\frac{4}{3(1+\bar{w}_{\text{reh}})}\ln 
{T_{\text{reh}}}
\ee
We represent the results graphically as parametric plots of $T_{\text{reh}}$ and $N_{\text{reh}}$ in 
 figure~{\ref{fig2}} using eq. \eqref{newN}, and figure~{\ref{fig3}} by using 
eq. \eqref{e38} and eq. 
\eqref{e39}. The figure~{\ref{fig2(a)}} represents  the variation 
of the reheat temperature at the end of 
reheating, $T_{\text{reh}}$, with reheating e-folds, $N_{\text{reh}}$, for six different values of the 
average 
equation of state parameter 
$\bar{w}_{\text{reh}}$. In figure~{\ref{fig2(b)}}, the reheating e-folds $N_{\text{reh}}$ 
as a function of the average 
equation of state parameter $\bar{w}_{\text{reh}}$ is shown for six different values of $T_{\text{reh}}$. 
From figure~{\ref{fig2}}, we see that for instantaneous reheating, $N_{\text{reh}}\rightarrow 0$, 
the 
temperature at the 
end of reheating is maximum and it is same for all values of $\bar{w}_{\text{reh}}$, which is the point 
where all 
curves converge in figure~{\ref{fig2(a)}}.\par
The relations of $T_{\text{reh}}$ and $N_{\text{reh}}$ to $n_{s}$, using eq. 
\eqref{e38} and eq. \eqref{e39}
are shown in 
figure~{\ref{fig3}}. Planck's central value of $A_{s}=2.139\times 10^{-9}$  and $ 
z_{\text{eq}}=3365$  are used and the parameter 
$\rho_{\text{eq}}$ is computed to be $10^{-9}$ GeV \cite{Ade:2015xua,Ade:2015lrj} to obtain the 
figure~{\ref{fig3}}.
 Within Planck's 1$\sigma$ bound on $n_{s}  (0.968\pm 0.006)$ \cite{Ade:2015xua} in  
figure~{\ref{fig3a}}, 
curves for 
 $-\frac{1}{3}\leq\bar{w}_{\text{reh}}\leq 0$  predict $T_{\text{reh}}>10^{7}$ GeV, while the other
curves ( $\frac{1}{6}\leq\bar{w}_{\text{reh}}\leq1$)  give all possible allowed  values of 
reheat temperature ($10^{-2}$ GeV to $10^{16}$ GeV). The $1\sigma$ and $2\sigma$ bounds on $n_{s}$ yield
$0.104\leq r \leq 0.152$ and $0.080\leq r \leq 0.176$ respectively  for $\frac{1}{2}m^{2}\phi^{2}$ potential. 
From figure~{\ref{fig3a}}, the corresponding $n_s$ range for
$\frac{1}{6}\leq\bar{w}_{\text{reh}}\leq1$ is $0.962 \lesssim n_s \lesssim 0.972$ which gives the upper and 
lower
bounds on tensor to scalar  ratio is $0.112\lesssim r \lesssim 0.152$.
 However, the upper bounds on the tensor to scalar ratio from recent observation is $r<0.09$ 
\cite{Ade:2015xua,Ade:2015lrj}.
 Hence, it is very difficult to find a feasible reheating scenario for $\frac{1}{2}m^2\phi^2$ inflaton 
 potential if the primordial gravitational waves are not  detected.\par
\subsection{Small field /Hilltop inflation:}
In this model, inflation occurs at very small value of the field and at the top of the flat 
potential. 
 The potential for this kind of inflation is described by \cite{Linde:1981mu,Kinney:1995ki,Martin:2013tda}.
\be\label{small}
V(\phi)= V_{0}\left[1-\left(\frac{\phi}{\mu}\right)^{p}\right] .
\ee
The field value at the end of inflation is calculated by setting $\epsilon=1$ and $\phi_{\text{end}}< \mu$ 
which leads 
to the following equation
\be 
\left(\frac{\phi_{\text{end}}}{\mu}\right)^{p}+\frac{p}{\sqrt{2}}\frac{M_{\text{Pl}}}{\mu}\left(\frac{\phi_{
\text{end}}}{\mu}\right)^{p-
1}=1 .
\ee
As per ref. \cite{Hazra:2010ve}, we have considered $p=4$ and $\mu=15 
M_{\text{Pl}}$ and obtained 
$\frac{\phi_{\text{end}}}{M_{\text{Pl}}}=14.34$.
For this quartic hilltop potential the Hubble parameter, defined in 
eq. \eqref{e21}, at the time of Hubble crossing of the 
scale $k$ takes the form 
\be\label{Hk}
H_{k}^{2}=\frac{1}{M_{\text{Pl}}^{2}}\left(\frac{V_{k}}{3-\epsilon_{k}}\right)=\frac{V_{0}}{M_{\text{Pl}}^{2}}
\left(\frac{
1-\left(\frac{\phi_{k}}{15M_{\text{Pl}}}\right)^{4}}{
3-\frac{1}{2}\left(\frac{-4\left(\frac{\phi_{k}}{M_{\text{Pl}}}\right)^{3}}{15^{4}-\left(\frac{\phi_{k}}{M_{
\text{Pl}}}
\right) ^{4}} \right)^{2} } \right).
\ee
As in the case of the large field quadratic model, we have to write the expressions for 
$T_{\text{reh}}$ 
and $N_{\text{reh}}$ as a function of $n_{s}$ and $A_{s}$. From eq. \eqref{e11} 
the expression for the  
number 
of e-folds between the horizon exit of the pivot scale 
and the end of inflation is given by
\be\label{e46}
\begin{aligned}
 \Delta 
N_{*}&=\frac{1}{M_{\text{Pl}}^{2}}\int_{\phi_{\text{end}}}^{\phi_{*}}{-\frac{\mu^{p}-\phi^{p}}{p\phi^{p-1}}
d\phi}\\
&=\frac{1}{M_{\text{Pl}}^{2}}\left[\frac{\mu^{p}}{p}\left(\frac{\phi_{\text{end}}^{2-p}}{2-p}-\frac{\phi_{*}^{
2-p}}{2-p}
\right)-\frac{1}{p}\left(\frac{\phi_{\text{end}}^{2}}{2}-\frac{\phi_{*}^{2}}{2}\right)\right ] .
 \end{aligned}
\ee
For $p=4$ and $\mu=15 M_{\text{Pl}}$ the eq. \eqref{e46} becomes 
\be\label{e46prime}
 \Delta 
N_{*}=6.328\times10^{3}\left[\left(\frac{M_{\text{Pl}}}{\phi_{*}}\right)^{2}-\left(\frac{1}{14.34}\right)^{2}
\right]
+\frac{1}{8}\left[\left(\frac{\phi_{*}}{M_{\text{Pl}}}\right)^{2}-(14.34)^{2}\right] .
\ee
We can write the field value at the time of horizon crossing of the pivot scale as a function of $n_{s}$ 
as given below
\be\label{e47}
\begin{aligned}
 n_{s}&=1-6\epsilon_{*}+2\eta_{*}\\
&=1-3M_\mathrm{Pl}^{2}\left(-\frac{4\phi_{*}^{3}}{(15M_{\mathrm
{
Pl } } )^ { 4 } -\phi_ {
* }^{4}}
\right)^{2}-\frac{24\phi_{*}^{2}}{(15M_{\mathrm{Pl}})^ { 4 } 
-\phi_{*}^{4}}M_{\mathrm{Pl}}^{2}
 \end{aligned}
\ee
\begin{figure}
\begin{subfigure}{.5\textwidth}
 \centering
  \includegraphics[width=1\linewidth]{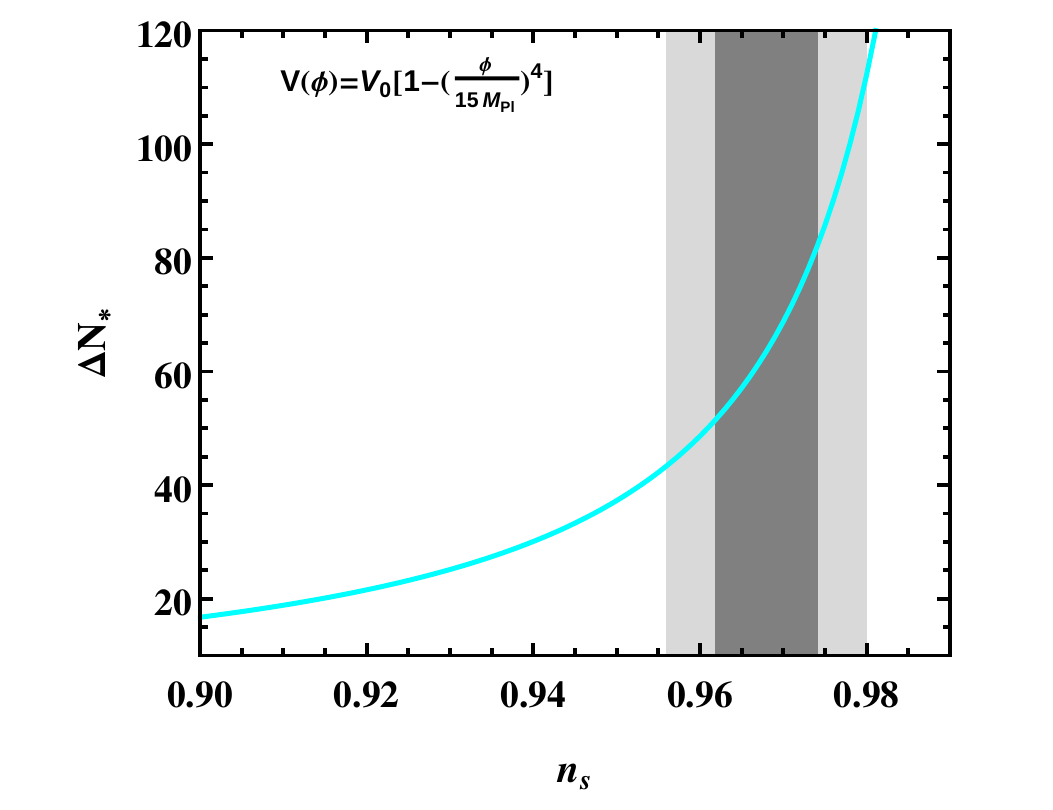}
  \caption{}
  \label{fig4a}
\end{subfigure}%
\begin{subfigure}{.5\textwidth}
  \centering
  \includegraphics[width=1\linewidth]{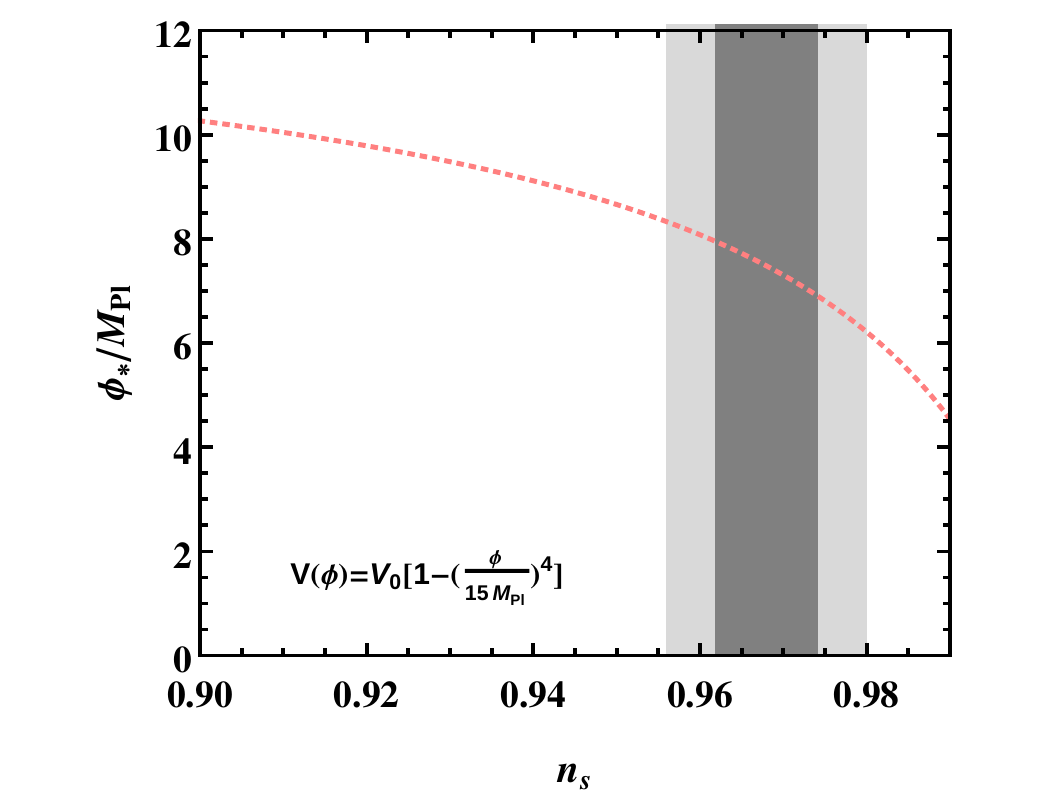}
  \caption{}
  \label{fig4b}
\end{subfigure}
\caption{(a) The number of e-folds remains ($\Delta N_*$) after the Hubble crossing of the pivot scale 
($k_{*}$) and (b) $\frac{\phi_{*}}{M_{\text{Pl}}}$  as a function of $n_{s}$ for small 
field/hilltop inflation with $p=4$ and $\mu=15 M_{\text{Pl}}$. Shading is same as figure~{\ref{fig1}}. }
\label{fig4}
\end{figure}
We can write $H_{*}$ and $V_{\text{end}}$ in terms of the scalar spectral amplitude $A_{s}$ and scalar 
spectral 
index $n_{s}$ as
\be
H_{*}^{2}&=&A_{s}8\pi^{2}M_{\text{Pl}}^{2}\epsilon_{*}=A_{s}4\pi^{2}M_{\text{Pl}}^{4}\left[\frac{-4\phi_{*}^{3
} } {
(15M_{\text{Pl}}^{4})-\phi_{
*}^{4}} \right]^{2},\\
 H_{*}&=&8\pi 
M_{\text{Pl}}\left(\frac{\chi^{3}(n_{s})}{15^{4}-\chi^{4}(n_{s})}\right)\sqrt{A_{s}}=8\pi 
M_{\text{Pl}}\beta(n_{s})\sqrt{A_{s}},\label{Hsmall}\\
\label{Vend}
 V_{\text{end}}&=&\gamma A_{s} 
M_{\text{Pl}}^{4}\frac{\beta^{3}(n_{s})\left(3-8\beta^{2}(n_{s})\right)}{\chi^{3}(n_{s})},
\ee
where, $\chi\tiny{(n_{s})}=\frac{\phi_{*}}{M_{\text{Pl}}}(n_{s})$ is the 
solution of eq. \eqref{e47} 
and is 
shown in figure~{\ref{fig4b}}. Here we define  
 $\beta(n_{s})=\frac{\chi^{3}(n_{s})}{15^{4}-\chi^{4}(n_{s})}$ and  $\gamma=5.28\times 
10^{6}$.
 Using above expressions for $H_{*}$ and $V_{\text{end}}$ we can write out $T_{\text{reh}}$ and 
$N_{\text{reh}}$ as a function of $n_{s}$ for the pivot scale but we shall not display the 
expressions here.
Finally the parameter $V_{0}$ entering in eq. \eqref{small} can be determined by 
using 
eqs. \eqref{Hk} and \eqref{Hsmall}. For hilltop potential with $p=4$ and 
$\mu=15 M_{\text{Pl}}$ using  Planck's central value of 
$n_{s}=0.968$  we get $V_{0}=(1.04\times 10^{16} 
\text{GeV})^{4}$ and $V_{\text{end}}=(6.64\times 10^{15} \text{GeV})^{4}$ which strongly places inflation in 
the Grand Unification epoch. After substituting the values of $V_{\text{end}}$ and $g_{\text{reh}}\approx 
100$ in eq. \eqref{e26} we obtain  the relation
\be
N_{\text{reh}}\approx\frac{142.64}{3(1+\bar{w}_{\text{reh}})}-\frac{4}{3(1+\bar{w}_{\text{reh}})}\ln 
T_{\text{reh}}.
\ee
.\par
\begin{figure}
\begin{subfigure}{.5\textwidth}
 \centering
  \includegraphics[width=1\linewidth]{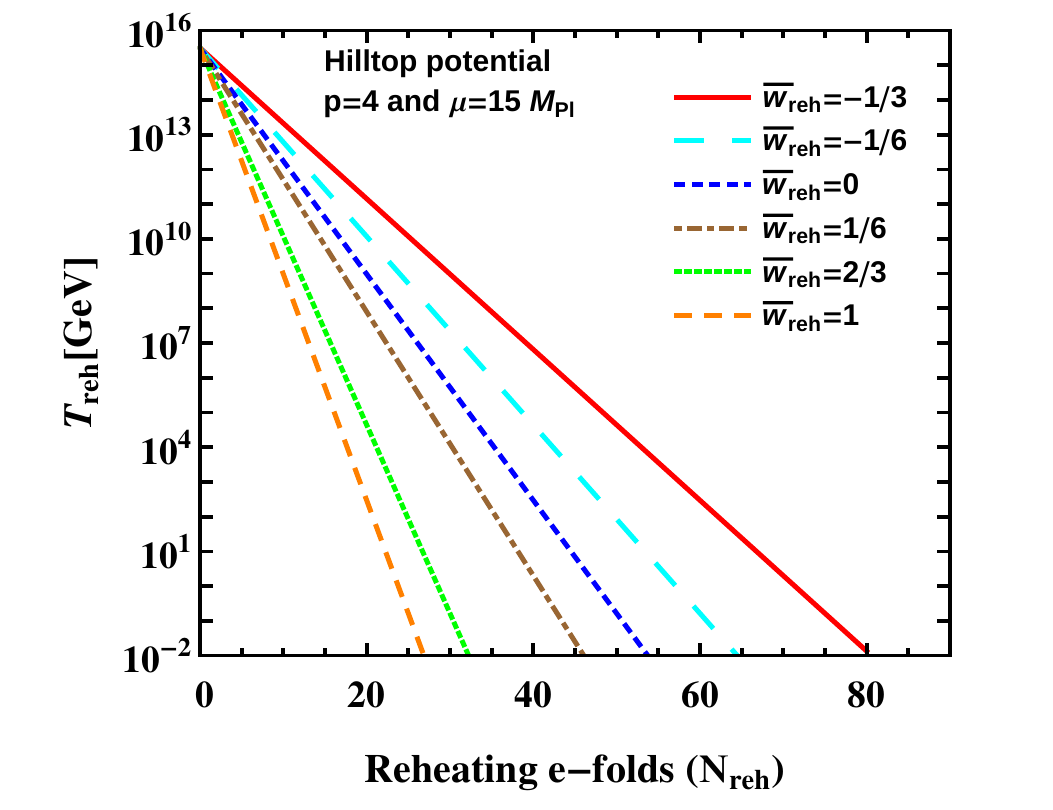}
  \caption{}
  \label{fig5a}
\end{subfigure}%
\begin{subfigure}{.5\textwidth}
  \centering
  \includegraphics[width=1\linewidth]{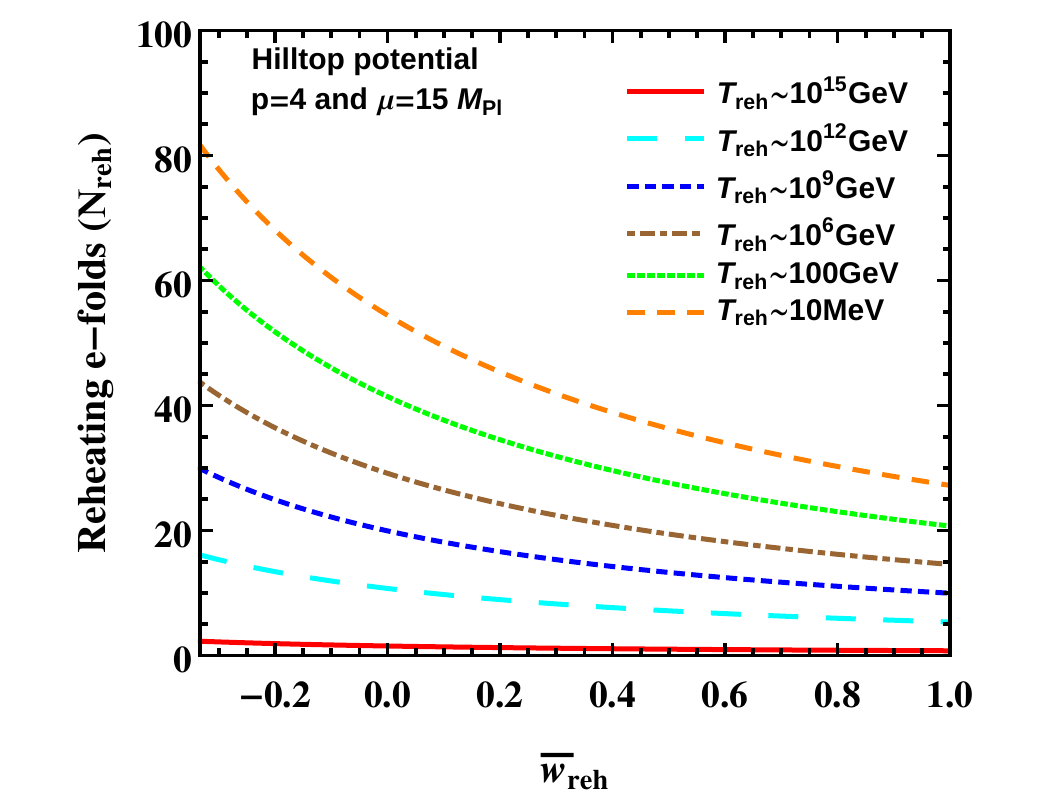}
  \caption{}
  \label{fig5b}
\end{subfigure}
\caption{Plots of  (a)  $T_{\text{reh}}$ with  
$N_{\text{reh}}$ and  (b)  
$N_{\text{reh}}$ as a function of $\bar{w}_{\text{reh}}$ for small field 
model with the same color coding as figure~{\ref{fig2}}.}
\label{fig5}
\end{figure}
\begin{figure}
\begin{subfigure}{.5\textwidth}
 \centering
  \includegraphics[width=1\linewidth]{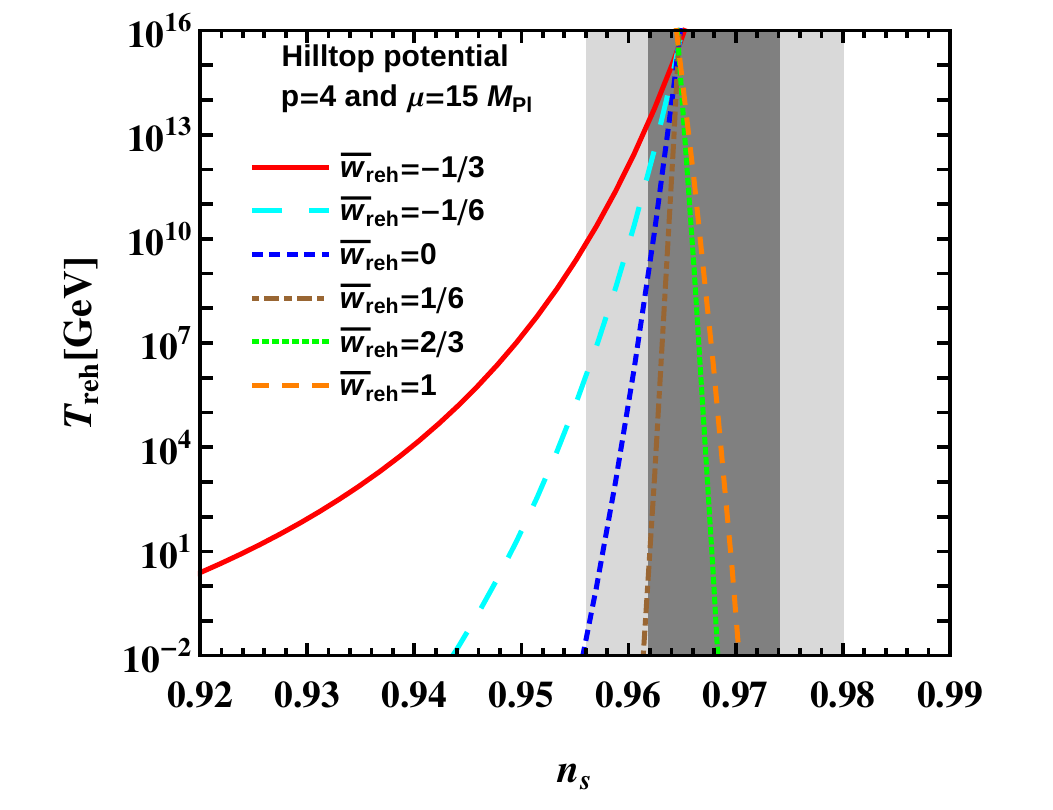}
  \caption{}
  \label{fig6a}
\end{subfigure}%
\begin{subfigure}{.5\textwidth}
  \centering
  \includegraphics[width=1\linewidth]{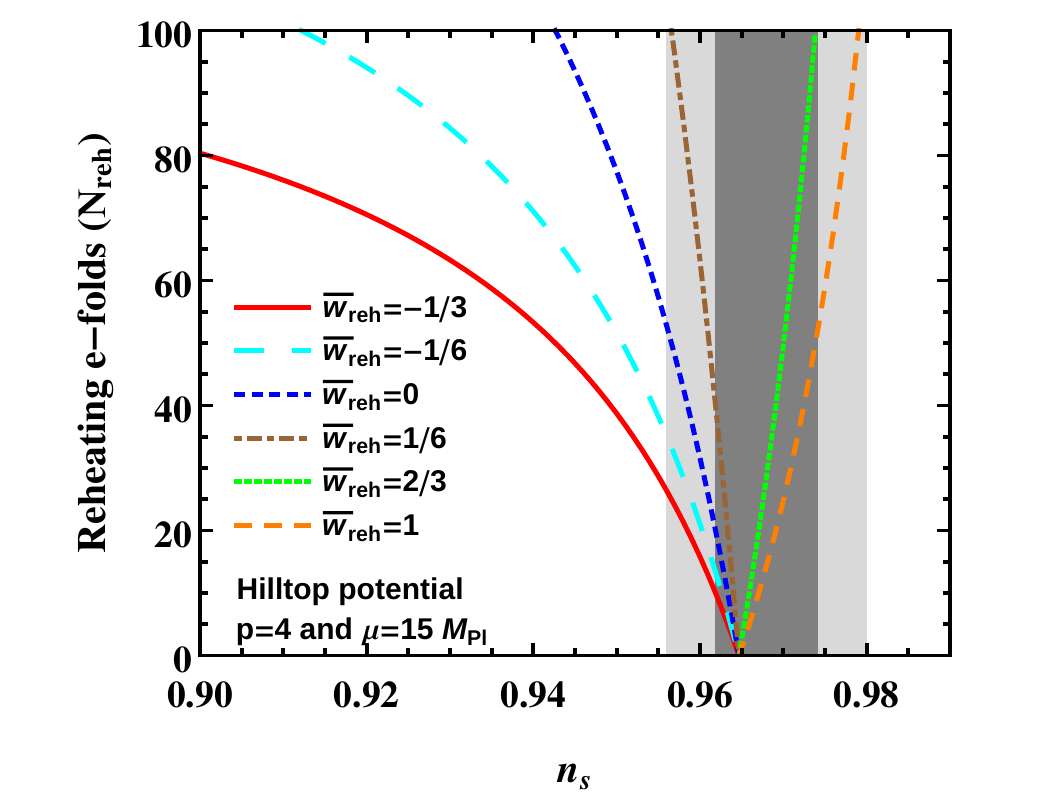}
  \caption{}
  \label{fig6b}
\end{subfigure}
\caption{Plots of $T_{\text{reh}}$ and $N_{\text{reh}}$ as a function of $n_{s}$, for small field model 
. All curves and shaded regions are as for figure~{\ref{fig3}}.}
\label{fig6}
\end{figure}

 The above results are represented graphically in figure~{\ref{fig5}} and figure~{\ref{fig6}}. 
The figure~{\ref{fig5a}} represents the  variation of $T_{\text{reh}}$ with $N_{\text{reh}}$ for 
different 
values of 
$\bar{w}_{\text{reh}}$ for quartic hilltop model. The length of reheating $N_{\text{reh}}$ as a function of  
$\bar{w}_{\text{reh}}$ is shown in figure~{\ref{fig5b}}.
 The variation of $T_{\text{reh}}$ and $N_{\text{reh}}$ as a 
function of $n_{s}$
are shown in figure~{\ref{fig6}} for 
different values of $\bar{w}_{\text{reh}}$. 
Within Planck's 1$\sigma $ bounds on $n_{s}$,  curves with $\bar{w}_{\text{reh}}<0$ 
( $\bar{w}_{\text{reh}}=-\frac{1}{3}$ and $ -\frac{1}{6}$ ) estimate 
$T_{\text{reh}}\gtrsim 10^{12}$ GeV as shown in figure~{\ref{fig6}}. 
The curves with $\bar{w}_{\text{reh}}\geq\frac{1}{6}$ span the entire range of reheating temperature i.e.,
$10^{-2}$ GeV to $10^{16}$ GeV within Planck's $1\sigma$ bounds on $n_{s}$.
For this quartic hilltop model  the tensor to scalar ratio r and the tensor tilt $n_{T}$ can be written as a 
function of 
$n_{s}$ and are  given below
\be\label{e54}
 r=16\epsilon=128\beta^{2}(n_{s}) ~~~~~\text{and} ~~~~~ n_{T}=-16\beta^{2}(n_{s})
\ee
 For the hilltop potential with $p=4$ and $\mu=15 M_{\text{Pl}}$, using eq. 
\eqref{e54} the $1\sigma$ and 
$2\sigma$ bounds 
 on $n_{s}$ yield $0.006\leq r \leq 0.015$ and $0.003\leq r \leq 0.020$ respectively.
 The constraint on the tensor  to scalar ratio $r\leq 0.02$ is obtained  from figure~{\ref{fig6a}} for 
$\bar{w}_{\text{reh}}\geq \frac{1}{6}$
 and $10^{-2}$ GeV $\leq T_{\text{reh}}\leq 10^{16}$ GeV within Planck's $1\sigma$ bounds on $n_{s}$.
   From figure~{\ref{fig6a}}, the upper bound $r\leq 0.015$ is estimated for 
 $\bar{w}_{\text{reh}}\geq 0$ and $10^{-2}$ GeV $\leq T_{\text{reh}}\leq 10^{16} $ GeV within
 Planck's $2\sigma$ bounds on $n_{s}$.\\
\subsection{Starobinsky Model:} 
The action for the Starobinsky model is 
\be
S=\frac{1}{2}\int{d^4x\sqrt{-g}\left(M_{\text{Pl}}^2R+\frac{1}{6M^2}R^2+\mathcal{L}_{\text{matter}}\right)},
\ee
where $R$ is the Ricci scalar and $\mathcal{L}_{\text{matter}}$ is the Lagrangian for the matter content of 
the Universe. After performing the conformal transformation \cite{Kalara:1990ar} to Einstein frame we get the 
action for the 
Starobinsky model which is equivalent to the scalar field version \cite{Kehagias:2013mya,Cook:2015vqa}, and 
is given below 
\be
S=\int{d^4x\Bigg[\frac{M_{\text{Pl}}^{2}}{2} 
R-\frac{1}{2}}\partial_{\mu}\phi\partial^{\mu}\phi-\frac{3}{4}M_{\text{Pl}}^2M^2\left(1-e^{-\sqrt{ \frac { 2 
} 
{3}}\phi/M_{\text{Pl}}} \right)^{2}\Bigg].
\ee
Hence, in Einstein frame the Lagrangian becomes normal gravity plus a scalar field $\phi$ with the potential 
\cite{Kehagias:2013mya} 
\be\label{StrP}
V(\phi)&=\frac{3}{4}M_{\text{Pl}}^{2}M^{2}\left(1-e^{-\sqrt{\frac { 2 } {3}}\phi/M_{\text{Pl}}} 
\right)^{2}\nn\\
&=\Lambda^{4}\left(1-e^{-\sqrt{\frac { 2 } {3}}\phi/M_{\text{Pl}}} \right)^{2}.
\ee
For the above potential given in eq. \eqref{StrP}, the slow-roll parameters take 
the form 
\be
\epsilon=\frac{4}{3}\frac{1}{\left(1-e^{\sqrt{\frac{2}{3}}\phi/M_{\text{Pl}}}\right)^{2}} \hspace{1cm} 
\text{and} \hspace{1cm} 
\eta=\frac{4}{3}\frac{\left(2e^{-\sqrt{\frac{2}{3}}\phi/M_{\text{Pl}}}-1\right)e^{-\sqrt{\frac{2}{3}}\phi/M_{
Pl } }}{\left(1-e^{-\sqrt{ \frac { 2 } 
{3}}\phi/M_{\text{Pl}}} \right)^{2}}.
\ee
The Hubble parameter as defined in eq. \eqref{e21} at the time of Hubble 
crossing of the scale $k$ is given 
by
\be\label{HkStr}
H_{k}^{2}=\frac{\Lambda^{4}}{M_{\text{Pl}}^{2}}\left(\frac{\left(1-e^{
-\sqrt{\frac { 2 } {3}}\phi_{k}/M_{\text
{ Pl } } } \right)^{2}}{ 
3-\frac{4}{3}\left(\frac{1}{1-e^{\sqrt{\frac{2}{3}}\phi_{k}/M_{\text{Pl}}}}\right)^{2}} \right).
\ee
The number of e-folds remaining after the  scale $k$ crosses the 
Hubble radius is obtained by using eq. \eqref{e11}, and is written below
\be\label{DeltaNStr}
\Delta 
N_{k}=\frac{3}{4}\left[e^{\sqrt{\frac{2}{3}}\phi_{k}/M_{Pl}}-\left(1+\frac{2}{\sqrt{3}}
\right)+\ln\left(1+\frac{2}{\sqrt{3}}\right)\right]-\sqrt{\frac{3}{8}}\frac{\phi_{k}}{M_{Pl}}.
\ee
 Now, similar to the large field and hilltop model we have to express 
$T_{\text{reh}}$ and $N_{\text{reh}}$ as a function of the experimentally estimated parameters $A_{s}$ and 
$n_{s}$.
The scalar spectral index $n_{s}$  as defined in eq. \eqref{ns} for this model 
becomes
\be\label{nsS}
n_{s}=1-\frac{8}{3}\frac{\left(1+e^{\sqrt{\frac{2}{3}}\phi_{*}/M_{\text{Pl}}}\right)
} { \left(1-e^ { \sqrt{\frac{2}{3}}\phi_{*}/M_{\text{Pl}}}\right)^{2}}.
\ee
From the above expression, eq. \eqref{nsS}, we can write the field value at the 
time of Hubble crossing 
of the pivot scale $k_{*}$ as a function of $n_{s}$ as 
\be
\frac{\phi_{*}}{M_{\text{Pl}}}=\sqrt{\frac{3}{2}}\ln\left(\frac{8}{3(1-n_{s})}\right) .
\ee
Now, we can write $\Delta N_{*}$, $H_{*}$ and $V_{\text{end}}$ as a function of the scalar spectral index 
$n_{s}$ as following 
\be
\Delta 
N_{*}=\frac{3}{4}\left[\frac{8}{3(1-n_{s})}-\left(1+\frac{2}{\sqrt{3}}\right)-\ln\left(\frac{8}{(1-n_{s}
)(3+2\sqrt{3}) } \right)\right],
\ee
\begin{figure}
\begin{subfigure}{.47\textwidth}
 \centering
  \includegraphics[width=1\linewidth]{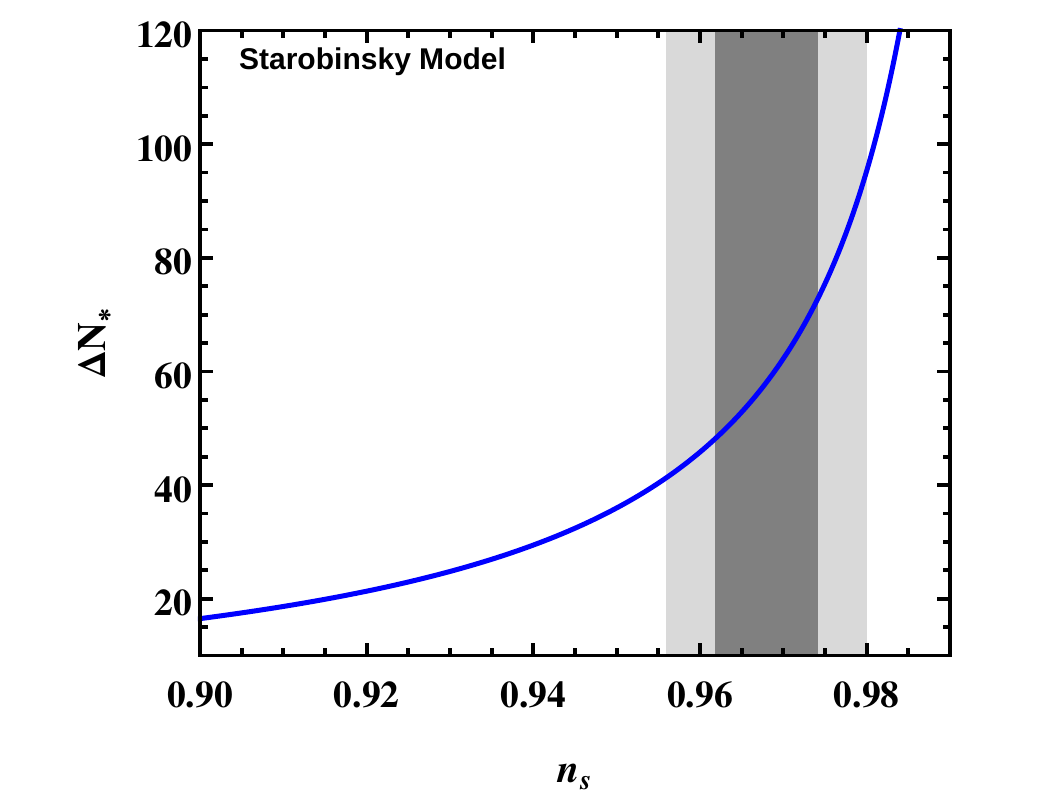}
  \caption{}
  \label{fig7a}
\end{subfigure}%
\begin{subfigure}{.47\textwidth}
  \centering
  \includegraphics[width=1\linewidth]{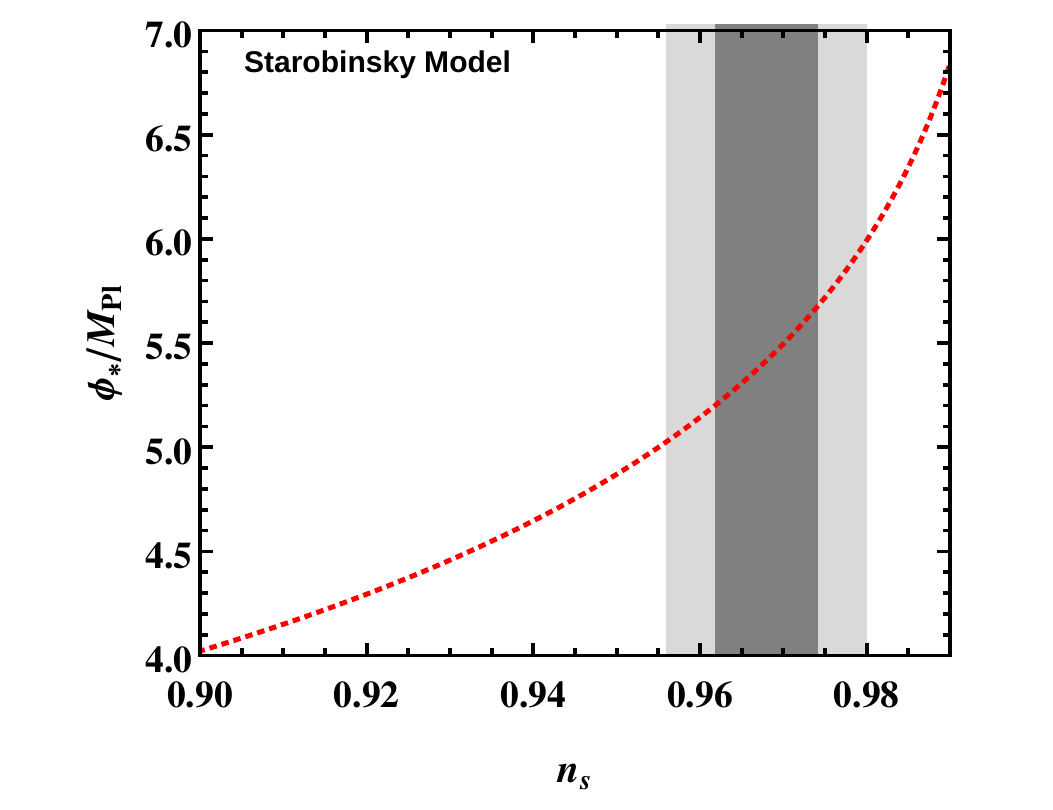}
  \caption{}
  \label{fig7b}
\end{subfigure}
\caption{The variation of (a)  $\Delta N_{*}$  and (b) $\frac{\phi_{*}}{M_{\text{Pl}}}$  as a function of 
scalar 
spectral index $n_{s}$ for Starobinsky model.
 Shading is same as figure~{\ref{fig1}}}
\label{fig7}
\end{figure}
\begin{figure}
\begin{center}
\begin{subfigure}{.5\textwidth}
  \includegraphics[width=1\linewidth]{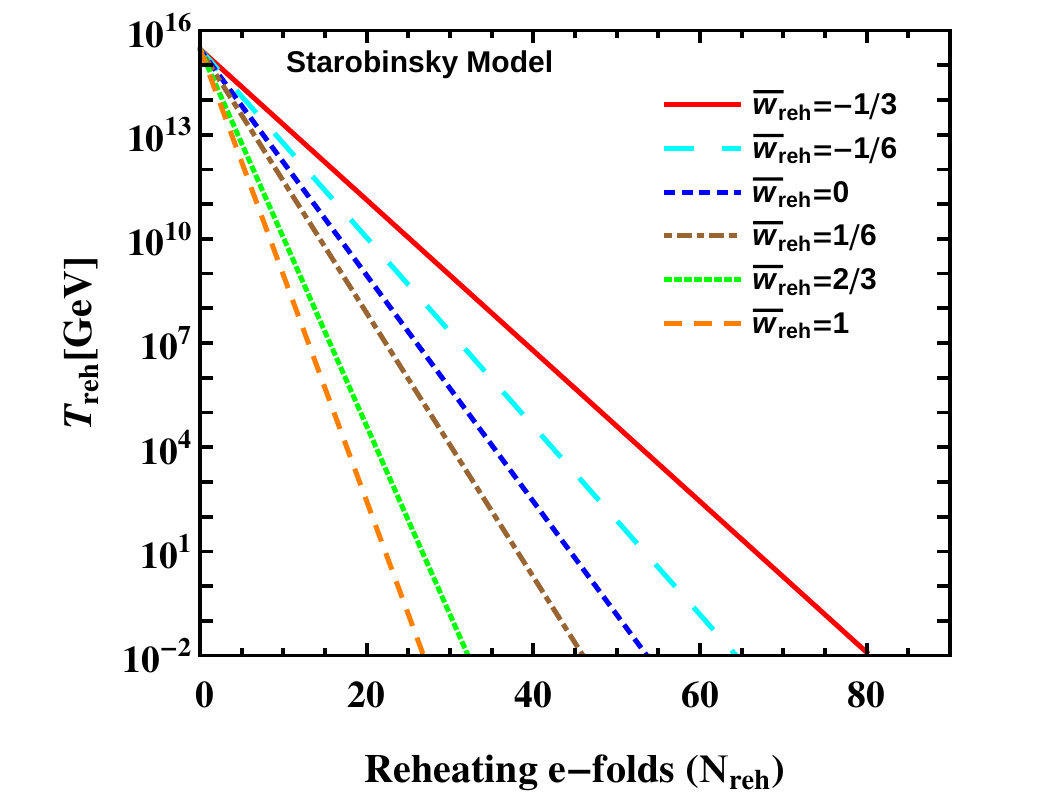}
  \caption{}
  \label{fig8(a)}
\end{subfigure}%
\begin{subfigure}{.5\textwidth}
  \includegraphics[width=1\linewidth]{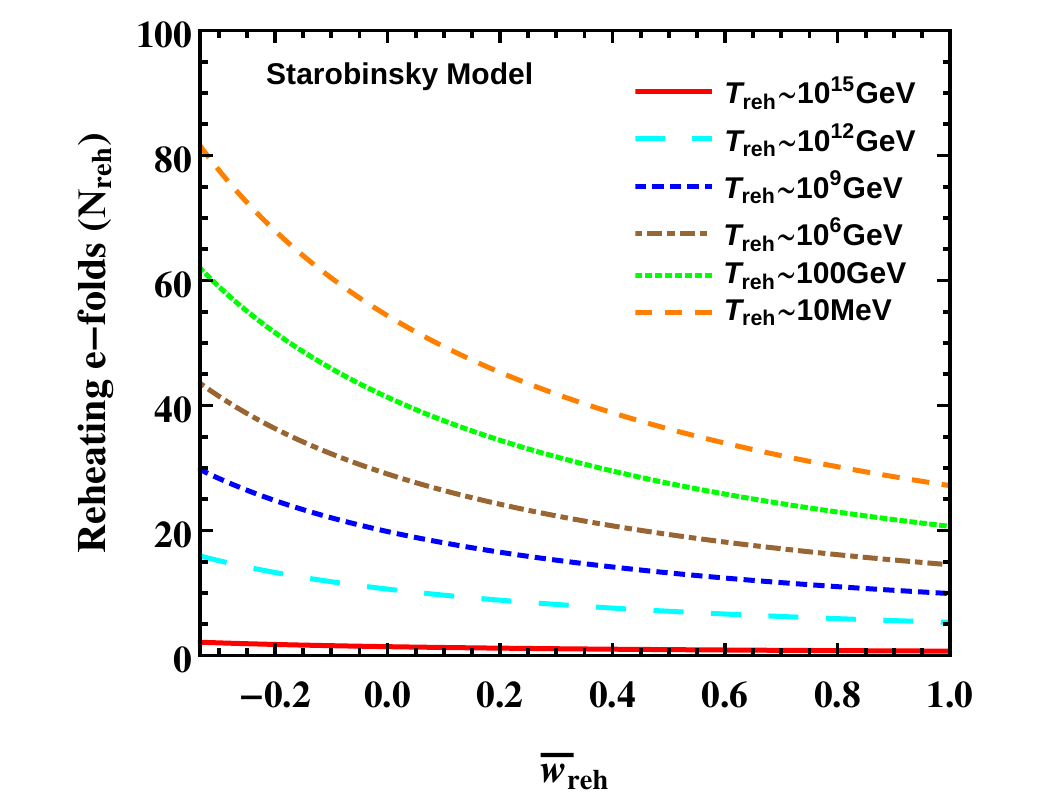}
  \caption{}
  \label{fig8(b)}
\end{subfigure}
\end{center}
\caption{Plots of  (a)  $T_{\text{reh}}$ with  
$N_{\text{reh}}$ and  (b)  
$N_{\text{reh}}$ as a function of $\bar{w}_{\text{reh}}$ for Starobinsky
model with the same color coding as figure~{\ref{fig2}}. }
\label{fig8}
\end{figure}
\begin{figure}
\begin{subfigure}{.5\textwidth}
 \centering
  \includegraphics[width=1\linewidth]{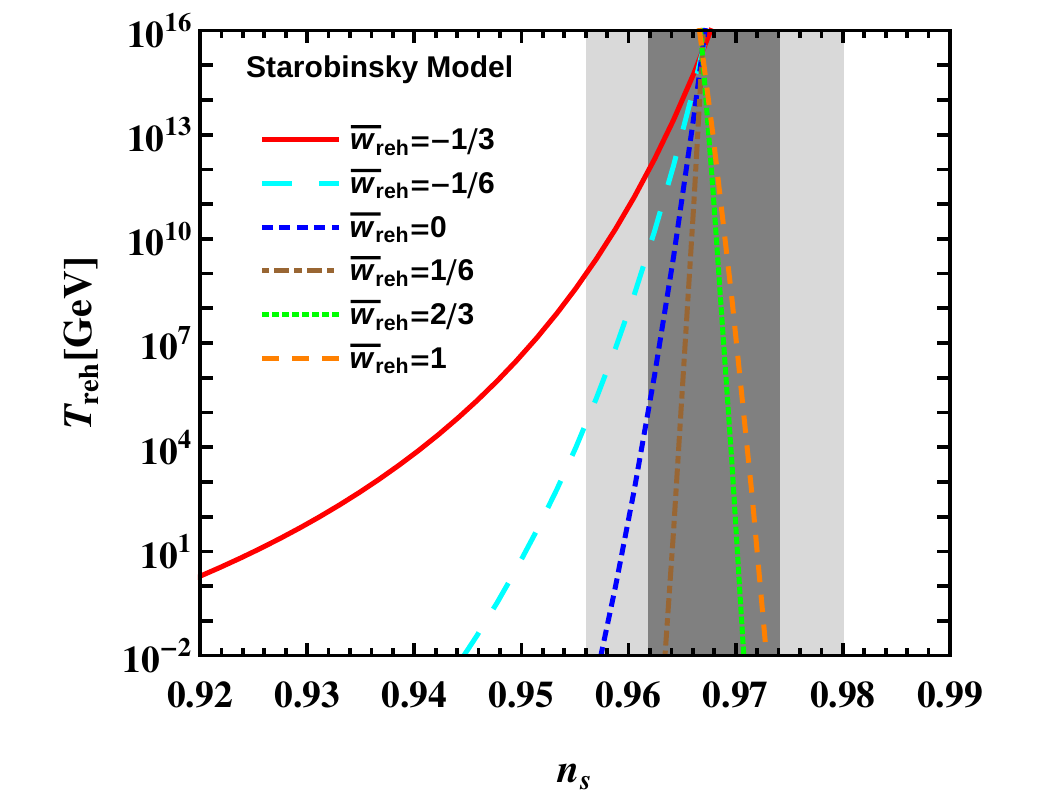}
  \caption{}
  \label{fig9a}
\end{subfigure}%
\begin{subfigure}{.5\textwidth}
  \centering
  \includegraphics[width=1\linewidth]{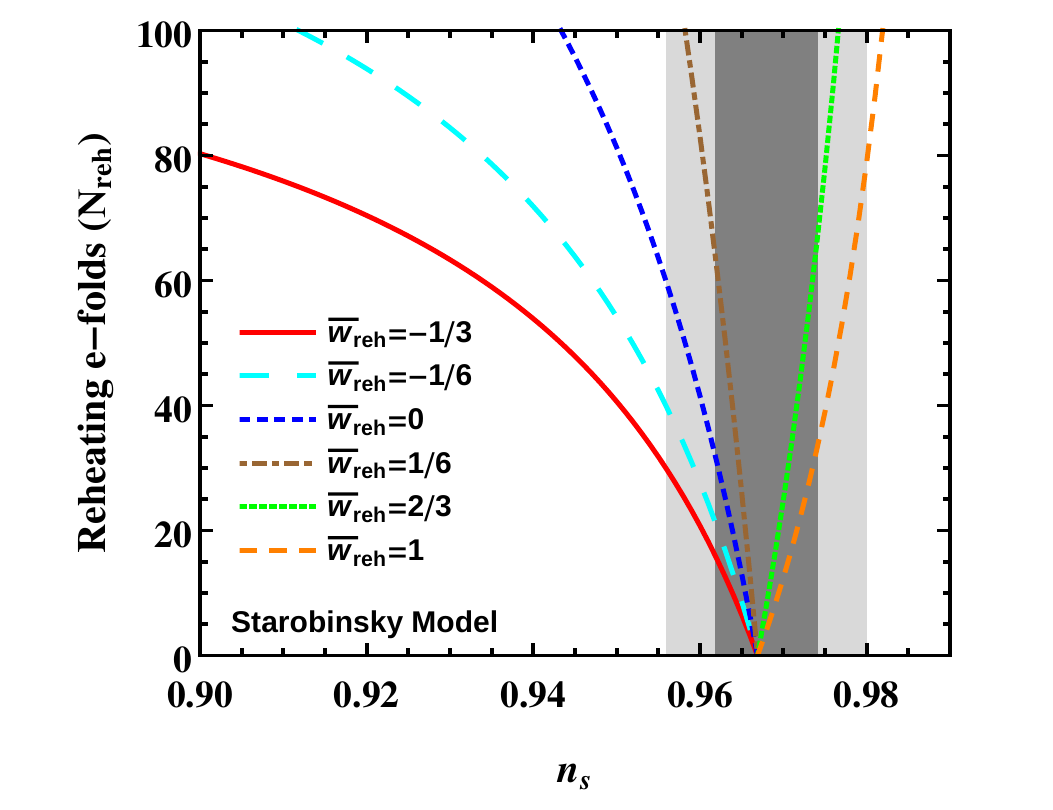}
  \caption{}
  \label{fig9b}
\end{subfigure}
\caption{Plots of $T_{\text{reh}}$ and $N_{\text{reh}}$ as a function of $n_{s}$, for Starobinsky model 
. All curves and shaded regions are as for figure~{\ref{fig3}}.}
\label{fig9}
\end{figure}
\be\label{StrH}
H_{*}\approx\pi M_{\text{Pl}}\left(1-n_{s}\right)\sqrt{\frac{3}{2}A_{s}} 
\ee
and
\be\label{StrV}
V_{\text{end}}\approx\frac{9}{2}\pi^{2}A_{s}M_{\text{Pl}}^{4}(1-n_{s})^{2}\frac{1}{\left(1+\frac{\sqrt{3}}{2}
\right)^{2}}.
\ee
Since $n_{s}\approx1$, we have 
set $[1-x(1-n_{s})]\approx 1$ in eqs. \eqref{StrH} and \eqref{StrV} where $x$ 
also remains 
less than unity in magnitude. After substituting the above expressions for 
$H_{*},~ \Delta N_{*} 
~\text{and} ~V_{\text{end}}$ in eqs. \eqref{e25} and \eqref{e27} we can write 
the expressions of 
$T_{\text{reh}}$ and $N_{\text{reh}}$ as a function of  $n_{s}$ for the pivot scale. The variation of $\Delta 
N_{*}$ and $\frac{\phi_{*}}{M_{\text{Pl}}}$ as a function of $n_{s}$ for this model are shown in 
figure~{\ref{fig7}}
By using Planck's central value of 
$A_{s}=2.139\times 10^{-9}$ and 
$n_{s}=0.968$  we get $\Lambda=(1.04\times 10^{16} 
\text{GeV})^{4}$ and $V_{\text{end}}=(6.26\times 10^{15} \text{GeV})^{4}$ for the Starobinsky model, where 
eqs. \eqref{HkStr} and \eqref{StrH} are employed. After substituting the values 
of 
$V_{\text{end}}$ and $g_{\text{reh}}$ in \eqref{e26}, we obtain the relation 
\be
N_{\text{reh}}=\frac{142.42}{3(1+\bar{w}_{\text{reh}})}-\frac{4}{3(1+\bar{w}_{\text{reh}})}\ln 
T_{\text{reh}}~.
\ee
The above results are represented graphically in figure~{\ref{fig8}} and figure~{\ref{fig9}}. From 
figure~{\ref{fig9}} we see that for $\frac{1}{6}\leq\bar{w}_{\text{reh}}\leq1$ the entire range of 
reheating temperature $10^{-2}\text{GeV}\lesssim T_{\text{reh}}\lesssim 10^{16}\text{GeV}$ compatible with 
BBN 
and inflation energy scales is allowed for Planck's $1\sigma$ limits on $n_{s}$. However, for 
$-\frac{1}{3}\leq\bar{w}_{\text{reh}}<\frac{1}{6}$ there is a lower bound on $T_{reh}$. For example, with 
$\bar{w}=-\frac{1}{3}$, $T_{\text{reh}}\gtrsim 10^{12}\text{GeV}$. For Starobinsky model the tensor to scalar 
ratio $r$ can be written in terms of the scalar 
spectral index $n_{s}$ as $r\approx 3(1-n_{s})^{2}$.
These ranges of $r$ are well inside the upper bound of $r$ from recent observations \cite{Ade:2015xua}.       
\section{Constraints on the reheating parameters from a step in the inflaton potential}\label{sec4}
In the previous section the expressions for $T_{\text{reh}}$ and $N_{\text{reh}}$ are written 
as a function of $\bar{w}_{\text{reh}}$, $n_{s}$ and $A_{s}$, see eqs. 
\eqref{e38} and \eqref{e39}. 
However, we can also write $T_{\text{reh}}$ and $N_{\text{reh}}$ in terms of 
the inflaton field location $\phi_{k}$  and the number of 
e-folds remaining, $\Delta N_{k}$, after a given scale $k$ crosses the Hubble radius. Consider the case of 
the 
quadratic large field model in which after substituting 
$V_{\text{end}}=\frac{1}{2}m^{2}\phi_{\text{end}}^{2}$ and eq. \eqref{Hubble} in 
eqs. \eqref{e25} and 
\eqref{e27} the expressions for $T_{\text{reh}}$ and $N_{\text{reh}}$ take the form 
\be\label{Tlarge}
 \ln \left(T_{\text{reh}}\right)=\frac{3(1+\bar{w}_{\text{reh}})}{3\bar{w}_{\text{reh}}-1}&\Bigg{\{}\ln 
\left(m\sqrt{\frac{\frac{1}{2}\left(\frac{\phi_{k}}{M_{\text{Pl}}}\right)^{2}}{3-2\left(\frac{
M_{\text{Pl}}}{\phi_{k}}\right)^{2}}}\right)-\ln 
\left(\frac{k}{a_{0}}\right)-\ln\left(1+z_{\text{eq}}\right)
-\Delta N_{k}\nn\\
&+\ln\left(\rho_{\text{eq}}^{\frac{1}{4}}\right)\Bigg{\}}-\frac{1}{3\bar{w}_{\text{reh}}-1)}\ln 
\left(\frac{3}{4}m^{2}\phi_{\text{end}}^{2}\right)
-\frac { 1 } { 4 } \ln\left(\frac { \pi^ { 2 } } { 30 } g_ { \text{reh} } \right).
\ee
\be\label{Nlarge}
N_{\text{reh}}=\frac{1}{3\bar{w}_{\text{reh}}-1}\ln\left(\frac{3}{4}m^{2}\phi_{\text{end}}^{2}\right)+\frac{4}
{3\bar{w}_{\text{reh}}-1}\Bigg{\{}&
\ln\left(\frac { k } { a_ { 0
}}\right)+\ln\left(1+z_{\text{eq}}\right)+\Delta N_{k} -\ln\left(\rho_{\text{eq}}^{\frac{1
}{4}}\right)\nn\\
& -\ln 
\left(m\sqrt{\frac{\frac{1}{2}\left(\frac{\phi_{k}}{M_{\text{Pl}}}\right)^{2}}{3-2\left(\frac{
M_{\text{Pl}}}{\phi_{k}}\right)^{2}}}\right)\Bigg{\}}
\ee
Similarly, we can write out the expressions of $T_{\text{reh}}$ and $N_{\text{reh}}$ for 
quartic hilltop and Starobinsky model but these are not displayed here.
If we know the field location 
$\phi_{k}$ for the mode with wavenumber $k$ then the $\Delta N_{k}$ can be 
computed by using eq. \eqref{e11}, 
which leads to express $T_{\text{reh}}$ and $N_{\text{reh}}$ as a function of $\bar{w}_{\text{reh}}$ only. 
Using this method we proceed to strengthen our previous section's results (i.e. the allowed ranges of 
$T_{\text{reh}}$, $N_{\text{reh}}$ and $\bar{w}_{\text{reh}}$) by considering an observable 
scale $k$ at current epoch where local anomalies have been observed in the CMB angular power spectrum.\par
The WMAP collaboration
\cite{Peiris:2003ff,Komatsu:2008hk,0067-0049-192-2-18,0067-0049-208-2-19} reported a dip and a bump near 
the multipoles $\ell=22$ and $\ell=40$  in the CMB angular power spectrum. The large scale power suppression, 
namely the dip near $\ell=22$, is also found in Planck 2013 and 2015 data 
\cite{Ade:2013zuv,Planck:2013jfk,Ade:2015xua,Ade:2015lrj}. There are studies 
\cite{Peiris:2003ff,Covi:2006ci,Hamann:2007pa,Mortonson:2009qv,Hazra:2010ve,Benetti:2012wu} which relate such 
anomalies to a feature in the otherwise monotonic inflaton potential. Such feature is shown to generate 
oscillations in the primordial power spectrum, and provide better fit to the CMB data 
near these multipole moments. The step is 
introduced by multiplying the inflaton potential by a function
\be
 V_{\text{step}}(\phi)=\left[1+y~\text{tanh}\left(\frac{\phi-\phi_{k}}{\Delta\phi}\right)\right],
\ee
where $y$, $\phi_{k}$ and $\Delta\phi$  are respectively the height, location 
and width of the step. The 
scales affected by the step are those which are crossing the horizon at the time the inflaton field reaches 
the step location. The details of this mechanism can be found in refs.
\cite{Adams:2001vc,Leach:2000yw,Peiris:2003ff,Covi:2006ci,Hamann:2007pa,Benetti:2012wu,
Mortonson:2009qv}. 
The locations of the step obtained in ref. \cite{Hazra:2010ve}
are $\frac{\phi_{k}}{M_{\text{Pl}}}=14.67$ for 
$V(\phi)=\frac{1}{2}m^{2}\phi^{2}$ model and $\frac{\phi_{k}}{M_{\text{Pl}}}=7.888$ for $ 
V(\phi)=V_{0}\left[1-\left(\frac{\phi}{\mu}\right)^{p}\right]$ model with $p=4$ and $\mu=15 M_{\text{Pl}}$.
The initial condition is considered such that the pivot 
scale
($\frac{k_*}{a_0}=0.05 \text{Mpc}^{-1}$) leaves the Hubble 
radius at 50 e-folds before the end of inflation. This gives us a handle on the time scale at 
which that particular mode left the horizon. Using eq. \eqref{NKE} we can thus 
determine the number of e-folds remaining after the Hubble crossing of the mode corresponding to $\ell=22$ to 
be $\Delta N_{k}\approx 53$. 
As discussed in sec. \ref{sec2},  $\Delta N_{k}$ receives a small correction from the brief event of getting past the
step in the potential. However, this correction remains small so long as the step itself is small, i.e. $y\ll 1$.
This is demonstrated for quadratic large field and  quartic hilltop potential  in appendix \ref{Ap}.  

For the Starobinsky model, 
considering $\Delta N_{k}\approx 53$ and using eq. 
\eqref{DeltaNStr} we obtain the field location $\frac{\phi_{k}}{M_{\text{Pl}}}\approx 5.33$. Now, using the 
step location, 
$\frac{\phi_{k}}{M_{\text{Pl}}}$, the Hubble parameter value $H_{k}$ at the time of 
Hubble crossing of the scale $k$ can be found to be  $H_{k}=8.27\times 10^{13}$ GeV, $2.46\times 
10^{13}$ GeV and $1.37\times 10^{13}$ GeV for $V(\phi)=\frac{1}{2}m^{2}\phi^{2}$ and
$V_{0}\left[1-\left(\frac{\phi}{15M_{\text{Pl}}}\right)^{4}\right]$ and Starobinsky model
respectively, where eqs. \eqref{Hubble}, \eqref{Hk} and \eqref{HkStr} are 
employed. \par
In order to use this result we need to relate the values of $\ell$ of the CMB power spectrum 
with the wave number of the fluctuation. A comoving scale $\lambda_{co}$ is projected on the last 
scattering surface on an angular scale $\theta$ is given by \cite{Riotto:2002yw}
\be\label{e68}
\theta=\frac{\lambda_{co}}{\int{_{t_{ls}}^{t_{0}}\frac{dt}{a(t)}}}=\frac{\lambda_{co}}{\int_{t_{ls}}
^{t_{0}}{\frac{dt}{a_{ls}[\frac{3}{2}H_{ls}(t-t_{ls})+1]^{2/3}}}}  ,
\ee
where, $t_{ls}$ and  $t_{0}$ are last scattering and present time respectively,
and with $a_{ls}$ and $H_{ls}$ the scale factor and Hubble parameter at the 
time of last scattering. Here we have 
considered the Universe is matter dominated $\left(a(t)\propto t^{2/3}\right)$ from the time of last 
scattering till 
today. This in turn allows expressing the multipole moment value $\ell$ as
\be\label{e72}
 \frac{2\pi}{\ell}\approx\theta=\frac{\lambda_{\text{ph}}H_{0}}{2}\frac{\sqrt{1+z_{ls}}}{\sqrt{1+z_{ls}}-1} ,
\ee
where $\lambda_{\text{ph}}$ is the physical wavelength
and $z_{ls}$ is the redshift of the last scattering which is defined as $1+z_{ls}=\frac{a_{0}}{a_{ls}}$.
Using eq. \eqref{e72} we can calculate the physical wavelength 
($\lambda_{\text{ph}}$) and wave-number 
($k_{\text{ph}}=\frac{2\pi}{\lambda_{\text{ph}}}$)  
 corresponding to the multipole moments $\ell$. The present values of $\lambda_{\text{ph}}$  and 
$k_{\text{ph}}$ corresponding
 to  $\ell=2,22$ and $40$ are shown in table{~\ref{tab1}}.\\
\begin{table}[H]
\centering
\begin{tabular}{ p{2cm}p{2.4cm}p{2.5cm}p{2.5cm}p{2.5cm}  }
 \hline\hline
Multipole moment&Physical wavelength (Mpc) &Physical wavelength 
($\text{GeV}^{-1}$)&Physical wavenumber 
($\text{Mpc}^{-1}$)&Physical wavenumber ($\text{GeV}$)\\
 \hline
 $\ell=2$ & $2.69\times 10^{4}$& $4.20\times 10^{42}$& $2.34\times 10^{-4}$ & $1.50\times 10^{-42}$ \\
 $\ell=22$&$2.45\times 10^{3}$& $3.82\times10^{41}$&$2.57\times 10^{-3}$& $1.65\times 10^{-41}$\\
 $\ell=40$&$1.35\times 10^{3}$ &$2.11\times 10^{41}$& $4.68\times 10^{-3}$& $3.00\times 10^{-41}$\\
 \hline\hline
\end{tabular}
\caption{Present value of the wavelengths and wavenumbers corresponding to multipole moments  $\ell=2,22$ and 
$40$.}
\label{tab1}
\end{table}
Using the value $\frac{k_{(\ell=22)}}{a_{0}}\sim 2.57\times10^{-3}\text{Mpc}^{-1}$ and 
$\Delta N_{k}\sim 53$ for the field location $\phi_{k}$ thus 
allows us to obtain relations between $T_{\text{reh}}$ and $\bar{w}_{\text{reh}}$ and $N_{\text{reh}}$ and 
$\bar{w}_{\text{reh}}$ directly. The mass term $m$ and the energy scales $V_{0}$ and 
$\Lambda$  appearing in these expressions are 
determined by using Planck's central value of $n_{s}=0.968$ and $A_{s}=2.139\times 10^{-9}$ 
\cite{Ade:2015xua,Ade:2015lrj}.\par
\begin{figure}
\begin{subfigure}{.5\textwidth}
 \centering
  \includegraphics[width=1\linewidth]{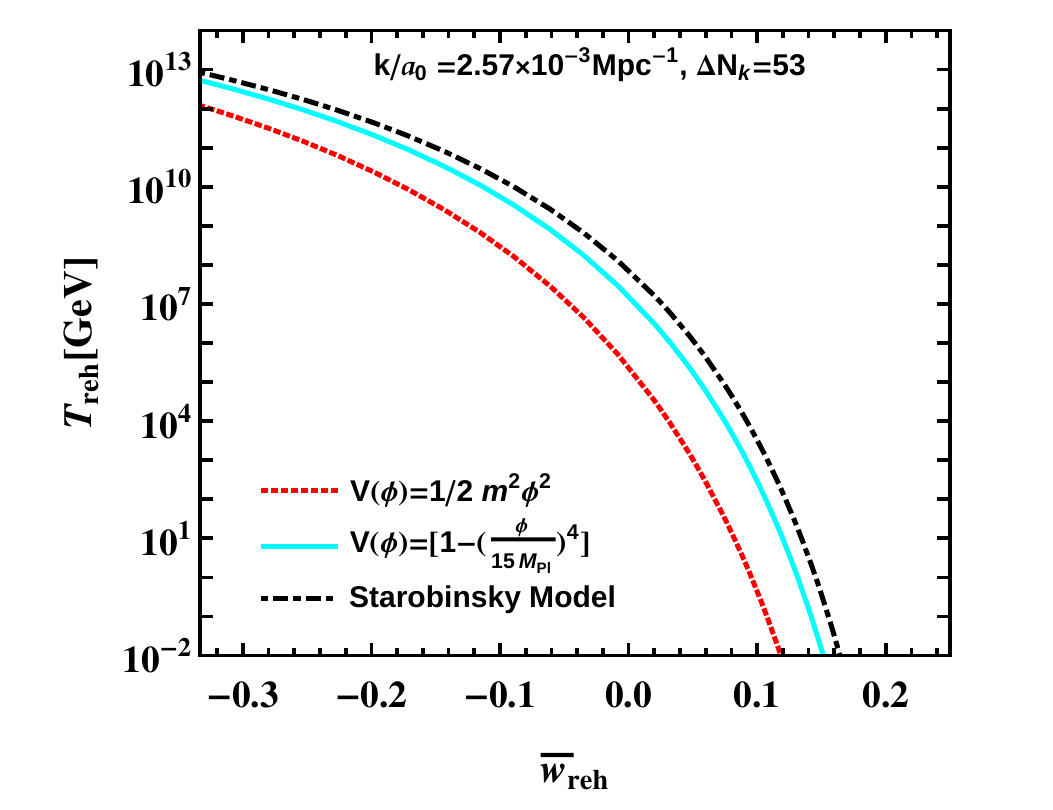}
  \caption{}
  \label{fig10a}
\end{subfigure}%
\begin{subfigure}{.5\textwidth}
  \centering
  \includegraphics[width=1\linewidth]{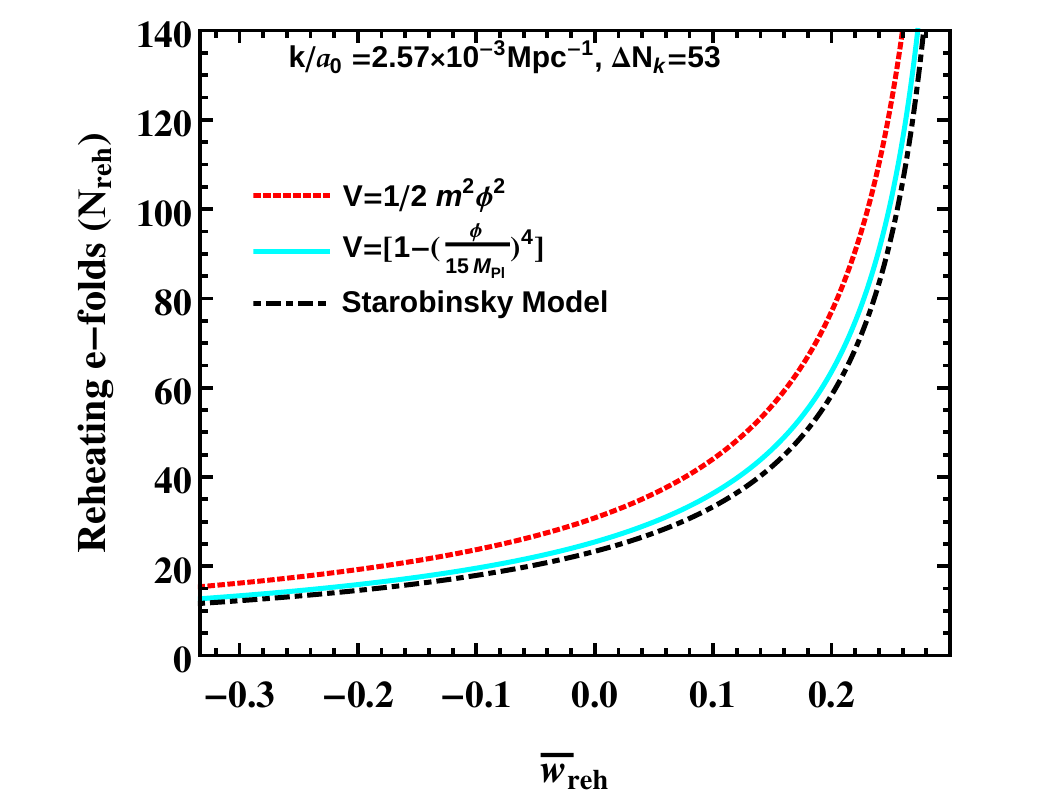}
  \caption{}
  \label{fig10b}
\end{subfigure}
\caption{Plots for $T_{\text{reh}}$ and $N_{\text{reh}}$ as a function of $\bar{w}_{\text{reh}}$ for 
different inflationary models. Tiny dashed red 
curve corresponds to $V=\frac{1}{2}m^{2}\phi^{2}$ model, dotdashed black curve represents the Starobinsky 
model
 and solid cyan curve is for hilltop potential with $p=4$ and $\mu=15 M_{\text{Pl}}$. 
}
\label{fig10}
\end{figure}

The resulting plots for $T_{\text{reh}}$ and $N_{\text{reh}}$ for different inflationary 
models are shown in figure~{\ref{fig10}}. For $V(\phi)=\frac{1}{2}m^{2}\phi^{2}$,  tiny dashed red 
curves are plotted in figure~{\ref{fig10a}} and figure~{\ref{fig10b}} by using 
eqs. \eqref{Tlarge} 
and \eqref{Nlarge}. The solid cyan 
curves in 
figure~{\ref{fig10a}} and figure~{\ref{fig10b}} represent 
$V(\phi)=V_{0}\left[1-\left(\frac{\phi}{15M_{\text{Pl}}}\right)^{4}\right]$. Dotdashed black curves are for 
the Starobinsky model. From figure~{\ref{fig10}}, it can be seen that in 
order to get the $T_{\text{reh}}$ within the reasonable 
range i.e., 
$10^{-2}$ GeV to $10^{16}$ GeV,  the  $\bar{w}_{\text{reh}}$ should be less than $0.2$ for all the 
models which are considered here. The allowed ranges of the reheating 
parameters for different inflationary models from Planck data and for successful explanation of 
the CMB low multipole anomalies are shown in 
table{~\ref{tab2}}.
\begin{table}
\centering
\begin{tabular}{ p{3.5cm}p{3cm}p{4.6cm}p{4.5cm} }
 \hline\hline
Inflationary\newline model&Reheating\newline parameter&Allowed range \newline from the Planck \newline 2015 
data
&Allowed range for successful explanation of CMB low multipole anomalies \\
 \hline\hline
 \multirow{3}{2.5cm}{Quadratic large field model}
 & $\bar{w}_{\text{reh}}$& 
$-\frac{1}{3}\le\bar{w}_{\text{reh}}\le1$ & $-\frac{1}{3}\le\bar{w}_{\text{reh}}\le 0.118$ \\
 & $T_{\text{reh}} (\text{GeV})$ & $10^{-2}\le T_{\text{reh}}\le 10^{16}$& $10^{-2}\le T_{\text{reh}}\le 
1.21\times 10^{12}$\\
& $N_{\text{reh}}$ & $0\le N_{\text{reh}}\le 74.42$& $15.41\le N_{\text{reh}}\le 
47.65$\\
\hline
 \multirow{3}{3cm}{{Hilltop model with $p=4$ and $\mu=15\text{Pl}$}}& $\bar{w}_{\text{reh}}$& 
$-\frac{1}{3}\le\bar{w}_{\text{reh}}\le1$ & $-\frac{1}{3}\le\bar{w}_{\text{reh}}\le 0.142$ \\
 & $T_{\text{reh}} (\text{GeV})$ & $10^{-2}\le T_{\text{reh}}\le 10^{16}$& $10^{-2}\le T_{\text{reh}}\le 
5.32\times
10^{12}$\\
& $N_{\text{reh}}$ & $0\le N_{\text{reh}}\le 102.01$& $12.72\le N_{\text{reh}}\le 
44.25$\\
\hline
 \multirow{3}{3cm}{{Starobinsky model}}& $\bar{w}_{\text{reh}}$& 
$-\frac{1}{3}\le\bar{w}_{\text{reh}}\le1$ & $-\frac{1}{3}\le\bar{w}_{\text{reh}}\le 0.165$ \\
 & $T_{\text{reh}} (\text{GeV})$ & $10^{-2}\le T_{\text{reh}}\le 10^{16}$& $10^{-2}\le T_{\text{reh}}\le 
8.48\times
10^{12}$\\
& $N_{\text{reh}}$ & $0\le N_{\text{reh}}\le 102.01$& $11.67\le N_{\text{reh}}\le 
46.12$\\
\hline\hline
\end{tabular}
\caption{The allowed ranges of the reheating parameters from Planck data and for successful explanation of 
the CMB low multipole anomalies due to the features in the inflaton potential 
\cite{Peiris:2003ff,Covi:2006ci,Hamann:2007pa,Mortonson:2009qv,Hazra:2010ve,Benetti:2012wu} .}
\label{tab2}
\end{table}
\section{Discussion and Conclusions}\label{sec5}
We have traced the evolution of the observable CMB scales 
from the time of Hubble crossing during inflation to 
present time. This is done by considering an intervening epoch of reheating whose duration is parametrized 
by $N_{\text{reh}}$ and which is characterized by an effective equation of state parameter 
$\bar{w}_{\text{reh}}$. It is possible to obtain the variation of $T_{\text{reh}}$ and $N_{\text{reh}}$ as a 
function of 
the scalar spectral index $n_{s}$ for single field inflationary models and allowing a wide range of values of 
the average equation of state parameter $\bar{w}_{\text{reh}}$. When this is done we find that for the 
quadratic large field and quartic hilltop models with $\bar{w}_{\text{reh}}\in\left[{-\frac{1}{3},0}\right]$, 
the obtained reheat temperature range is approximately $T_{\text{reh}}\in\left[10^{6},10^{16}\right]$ GeV. If 
the reheat temperature is allowed to vary over the whole range 
$T_{\text{reh}}\in\left[10^{-2},10^{16}\right]$ GeV compatible with BBN the $\bar{w}_{\text{reh}}$ is 
restricted to fall within $\left[{\frac{1}{6},1}\right]$ compatible with the $1\sigma$ bounds on $n_{s}$. 
The same $T_{\text{reh}}$ range remains allowed to the Starobinsky model, while the range of the 
$\bar{w}_{\text{reh}}$ 
is a little wider i.e., $\bar{w}_{\text{reh}}\in\left[\approx{\frac{1}{12},1}\right]$.\par
However, modeling the CMB low multipole anomalies through  feature in the inflaton potential 
gives a further 
handle on the reheat parameters. The obtained upper 
bounds on the reheating parameters for the large field, hilltop and Starobinsky model are 
$(\bar{w}_{\text{reh}}<0.118$, $T_{\text{reh}}<1.21\times 10^{12}\text{GeV})$, 
$(\bar{w}_{\text{reh}}<0.142$, $T_{\text{reh}}<5.32\times10^{12}\text{GeV})$ and 
$(\bar{w}_{\text{reh}}<0.165$, $T_{\text{reh}}<8.48\times10^{12}\text{GeV})$
 respectively. However, in quadratic large field model for $\bar{w}_{\text{reh}}\leq 1$ implies $r\geq 
0.112$ which is found to be
greater than the recently observed upper bound on $r$ (i.e., $r<0.09$). This further strains the validity of
the quadratic large field model. On the other hand, for the hilltop and Starobinsky 
models in the allowed $\bar{w}_{\text{reh}}$ range, the tensor to 
scalar ratio $r$ is always well inside the upper bounds on $r$. Hence, these 
models are strongly 
preferred for the purpose of inflationary model building. In conclusion, if the low multipole 
anomalies are to be successfully explained by considering a step in the inflaton potential it is 
possible to put stronger constraints on the reheating parameters.
\section*{Acknowledgement}
RG would like to appreciate the help of Mr. Sangem Rajesh.
\appendix
\section{Computation of $\Delta N_{k}$ for quadratic large field and quartic hilltop potential 
with a step.}\label{Ap}
For the quadratic large field model the potential with the step is given by
\be
V(\phi)=\frac{1}{2}m^{2}\phi^{2}\left[1+y\tanh\left(\frac{\phi-\phi_{k}}{\Delta\phi}\right)\right],
\ee
where $y,\phi_{k} ~\text{and}~ \Delta\phi$ are respectively the height, location and width of the step. The 
derivative of the potential with respect to the field $\phi$ becomes 
\be\label{A2}
V'(\phi)=m^{2}\phi\left[1+y\tanh\left(\frac{\phi-\phi_{k}}{\Delta\phi}\right)\right]+\frac{1}{2}m^{2}
\phi^{2}\frac{y}{\Delta\phi}\text{sech}^{2}\left(\frac{\phi-\phi_{k}}{\Delta\phi}\right).
\ee
The change in the potential energy due to the step, 
$\Delta V=2yV(\phi_{k})$,
gives an additional slope
\be
\frac{\Delta V}{\Delta\phi}=\frac{2yV(\phi_{k})}{\Delta\phi}.
\ee
Which should be smaller than the actual slope, i.e., $\left|\frac{\Delta 
V}{\Delta\phi}\right|<|V'|$, hence 
\be\label{ANEW}
\left|\frac{m^{2}\phi_{k}^{2}y}{\Delta\phi}\right|<\left| m^{2}\phi_{k}\right|
\Rightarrow\left|\frac{y}{\Delta \phi}\right|< \left|\frac{1}{\phi_{k}}\right| .
\ee
For the quadratic large field model with $y=16.06\times 10^{-4},~\phi_{k}=14.67~M_{\text{Pl}}
~\text{and}~\Delta\phi=0.0311~M_{\text{Pl}} $ \cite{Hazra:2010ve} the above 
condition \eqref{ANEW} is satisfied. 
Hence we can treat the step as a small perturbation on the potential.\newline
The equation of motion of the inflaton field is 
\be\label{A3}
\ddot{\phi}+3H\dot{\phi}+V'(\phi)=0.
\ee
After substituting eq. \eqref{A2} in eq. \eqref{A3} we can write the 
acceleration of the inflaton field as  
\be\label{A4}
\ddot{\phi}=-3H\dot{\phi}-m^{2}\phi\left[1+y\tanh\left(\frac{\phi-\phi_{k}}{\Delta\phi}\right)\right]
-\frac{1}{2}m^{2}
\phi^{2}\frac{y}{\Delta\phi}\text{sech}^{2}\left(\frac{\phi-\phi_{k}}{\Delta\phi}\right).
\ee
Now, let $\dot{\phi}_{bs}$ is the velocity of the inflaton before the step. If  $\dot{\phi}$ is the 
velocity of the field when it crossing the step at a distance  $\delta\phi$ from $\phi_{bs}$ then we 
can write 
\be\label{A5}
\dot{\phi}^{2}-\dot{\phi}_{bs}^{2}&=&2\ddot{\phi}\delta\phi\nn\\
&=&-6H\dot{\phi}\delta\phi-2m^{2}\phi\left[1+y\tanh\left(\frac{\phi-\phi_{k}}{\Delta\phi}\right)\right]
\delta\phi\nn\\
&&-
m^{2}\phi^{2}\frac{y\delta\phi}{\Delta\phi}\text{sech}^{2}\left(\frac{\phi-\phi_{k}}{\Delta\phi}\right).
\ee
 However, The kinetic energy 
of the field can be written in terms of the slow-roll parameter even in the 
absence of slow roll as ( recalling eq. \eqref{kinetic} which is an exact 
equation) 
\be\label{A6}
\frac{\dot{\phi}^{2}}{2}=\frac{\epsilon V(\phi)}{3-\epsilon}.
\ee
Thus we can approximate the change in kinetic energy as
\be\label{A7}
\frac{1}{2}\left(\dot{\phi}^{2}-\dot{\phi}_{bs}^{2}\right)=\left(\frac{\epsilon 
V(\phi)}{3-\epsilon}-
\frac{\epsilon_{bs} V(\phi_{bs})}{3-\epsilon_{bs}}\right)\approx \frac{\Delta\epsilon V(\phi)}{3-\epsilon}.
\ee
where we ignore corrections proportional to the parameter $y$,  
Substituting eq. \eqref{A7} in eq. \eqref{A5} we obtain
\be
\frac{2\Delta\epsilon 
V(\phi)}{3-\epsilon}&=&-6H\dot{\phi}\delta\phi-2m^{2}\phi\left[1+y\tanh\left(\frac{\phi-\phi_{k}}{
\Delta\phi }\right)\right]
\delta\phi\nn\\
&&-
m^{2}\phi^{2}\frac{y\delta\phi}{\Delta\phi}\text{sech}^{2}\left(\frac{\phi-\phi_{k}}{\Delta\phi}\right).
\ee
Which gives 
\be\label{A9}
\dot{\phi}&=&-\frac{1}{3H\delta\phi}\Bigg\{ \frac{\Delta\epsilon 
V(\phi)}{3-\epsilon}+m^{2}\phi\left[1+y\tanh\left(\frac{\phi-\phi_{k}}{
\Delta\phi }\right)\right]
\delta\phi\nn\\
&&+\frac{1}{2}
m^{2}\phi^{2}\frac{y\delta\phi}{\Delta\phi}\text{sech}^{2}\left(\frac{\phi-\phi_{k}}{\Delta\phi}
\right)\Bigg\}.
\ee
The number of e-folds remaining after the field reaches the position $\phi_{k}$ is 
\be\label{A10}
\Delta 
N_{k}&=&\int_{\phi_{k}}^{\phi_{e}}\frac{H}{\dot{\phi}}d\phi=\int_{\phi_{k}}^{
\phi_{k}-\Delta\phi}
\frac 
{H}{\dot{\phi}}d\phi+\int_{\phi_{k}-\Delta\phi}^{\phi_{e}}
\frac { H } { \dot{\phi}}d\phi\nn\\
&\approx&\frac{1}{M_{\text{Pl}}^{2}}\int_{\phi_{e}}^{\phi_{k}-\Delta\phi}\frac{V
}{V'}d\phi+
\int_{\phi_{k}-\Delta\phi}^{\phi_{k}}\left(-\frac{H}{\dot{\phi}}
\right)d\phi\nn\\
&=&\Delta N_{\text{SR}}+\Delta N_{\text{step}}.
\ee
 Here the subscript  `SR' denotes evaluation of the quantity by considering 
the slow-roll 
approximation. For 
this quadratic large field potential, $\phi_{e}=\sqrt{2}M_{\text{Pl}}, 
\phi_{k}=14.67M_{\text{Pl}}
~\text{and}~\Delta\phi=0.0311M_{\text{Pl}}$ \cite{Hazra:2010ve}. As a result we 
obtain
\be\label{A11}
\Delta 
N_{\text{SR}}\approx\int_{\sqrt{2}}^{14.67-0.0311}\frac{V}{V'}d\phi \approx
53.07~.
\ee
Now we compute $\Delta N_{\text{step}}$  as presented below
\be\label{A12}
\Delta 
N_{\text{step}}=\int_{\phi_{k}-\Delta\phi}^{\phi_{k}}\left(-\frac{H}{
\dot{\phi}}
\right)d\phi.
\ee
After substituting the expression of $\dot{\phi}$, eq. \eqref{A9} in eq. 
\eqref{A12}
\be
&\Delta 
N_{\text{step}}=\int_{\phi_{k}-\Delta\phi}^{\phi_{k}}{
\frac
{ 3H^{2 } \delta\phi}{\frac{
\Delta\epsilon 
V(\phi)}{3-\epsilon}+m^{2}\phi\left[1+y\tanh\left(\frac{\phi-\phi_{k}}{
\Delta\phi }\right)\right]\delta\phi+\frac{1}{2}
m^{2}\phi^{2}\frac{y\delta\phi}{\Delta\phi}\text{sech}^{2}\left(\frac{\phi-\phi_{k}}{\Delta\phi}
\right)}}d\phi.\nn\\
\ee
\be
\approx \frac{3H_{k}^{2}\Delta\phi^{2}}{\frac{
\Delta\epsilon 
V(\phi_{k})}{3-\epsilon_{k}}+m^{2}\phi_{k}\Delta\phi+\frac{1}{2}
m^{2}\phi_{k}^{2}y}
\ee
where we have estimated the integration over this short interval by the value of 
the integrand at $\phi_{k}$ times the width of the interval. 
The numerical value of $\Delta N_{\text{step}}$ is obtained by substituting the 
values of 
 $\epsilon_{k}\approx 10^{-4},~ \Delta\epsilon\approx 10^{-4},~ 
y=16.06\times 10^{-4},~\phi_{k}=14.67~M_{\text{Pl}}
~\text{and}~\Delta\phi_{k}=0.0031~M_{\text{Pl}}$ \cite{Hazra:2010ve} and, is 
given by
\be\label{A15}
\Delta N_{\text{step}}\approx 0.16~.
\ee
Finally, for the large field model we obtain
\be\label{A16}
\Delta N_{k}=\Delta N_{\text{SR}}+\Delta 
N_{\text{step}}\approx 53.07+0.16=53.23~.
\ee
Now, in the case of quartic hilltop model the field is rolling from lower value 
to it's higher value in the potential, hence we can write 
\be
\Delta 
N_{k}&=&\int_{\phi_{k}}^{\phi_{e}}\frac{H}{\dot{\phi}}d\phi=\int_{\phi_{k}}^{
\phi_{k}+\Delta\phi}
\frac 
{H}{\dot{\phi}}d\phi+\int_{\phi_{k}+\Delta\phi}^{\phi_{e}}\frac{H}{
\dot{\phi}}d\phi\nn\\
&\approx&\frac{1}{M_{\text{Pl}}^{2}}\int_{\phi_{e}}^{\phi_{k}+\Delta\phi
}\frac{V}{V'}d\phi+
\int_{\phi_{k}+\Delta\phi}^{\phi_{k}}\left(-\frac{H}{\dot{\phi}}
\right)d\phi\nn\\
&=&\Delta N_{\text{SR}}+\Delta N_{\text{step}}.
\ee
For this quartic hilltop model  we have $\phi_{e}=14.34 M_{\text{Pl}}$, 
$\phi_{k}=7.888$ and 
$\Delta\phi=0.0090M_{\text{Pl}}$ which gives
\be\label{B7}
\Delta 
N_{\text{SR}}\approx\int_{14.34}^{7.888+0.0090}\frac{V}{V'}
d\phi\approx52.79~.
\ee
Now, Similar to the large field model we can write the expression of $\Delta 
N_{\text{step}}$ for this hilltop model, and is 
\be
&\Delta 
N_{\text{step}}=\int_{\phi_{k}+\Delta\phi}^{\phi_{k}}\frac{ 
3H^ { 2 }
\delta\phi}{\frac{
\Delta\epsilon 
V(\phi)}{3-\epsilon}+\frac{V_{0}\phi^{3}\delta\phi}{\mu^{4}}\left[
1+y\tanh\left(\frac { \phi-\phi_{k}}{
\Delta\phi}\right)\right]-V_{0}\left[1-\left(\frac{\phi}{\mu}\right)^{4}\right]
\frac{y\delta\phi}{
\Delta\phi}
\text { sech}^{2}
\left(\frac{\phi-\phi_{k}}{\Delta\phi}\right)}d\phi\nn\\
\ee
\be\label{B10}
\approx \frac{3 H_{k}^{2}{\Delta\phi}^{2}}{\frac{
\Delta\epsilon 
V(\phi_{k})}{3-\epsilon_{k}}+\frac{V_{0}\phi_{k}^{3}\Delta\phi}{\mu^{4}}-V_{0}
y\left[1-\left(\frac{\phi_
{ k } } { \mu } \right)^ { 4 }
\right]}~.
\ee
Substituting $\epsilon_{k}\approx10^{-4},\Delta\epsilon\approx 10^{-4}, 
\phi_{k}=7.888M_{\text{Pl}}, \Delta\phi=0.0090 M_{\text{Pl}}$ and 
$y=-0.1569\times10^{-3}$ 
\cite{Hazra:2010ve} in eq. 
\eqref{B10} we obtain
\be\label{B11}
\Delta N_{\text{step}}\approx 0.28~.
\ee
Adding eqs. \eqref{B7} and \eqref{B11} we get 
\be\label{B12}
\Delta N_{k}=\Delta N_{\text{SR}}+\Delta 
N_{\text{step}}\approx 52.79+0.28=53.07~.
\ee
Hence from the above results, i.e., \eqref{A16} and \eqref{B12} we can consider 
that 
$\Delta N_{k}\approx 53$.
\bibliographystyle{JHEP}
\bibliography{Inflation}
\end{document}